\documentclass[aps,prc,showpacs,showkeys,superscriptaddress,nofootinbib,twocolumn,floatfix]{revtex4}
\usepackage{placeins}
\usepackage[]{graphicx}
\usepackage{float}
\usepackage{multirow}
\usepackage{color}
\usepackage{subfigure}
\usepackage{amsmath}
\usepackage{amssymb}
\usepackage{mathtools}
\usepackage{braket}
\usepackage[compat=1.1.0]{tikz-feynman}
\usepackage{graphicx,color,amsmath}
\usepackage[colorlinks=true, linkcolor=blue, citecolor=blue, urlcolor=blue]{hyperref}

\begin{document}

\title{
The Connection between the Stochastic Schr\"{o}dinger Equation and Boltzmann Equation}
\author{Zichao Li}
\affiliation{
  School of Nuclear Science and Technology, University of Chinese Academy of Sciences, Beijing 100049, China
}
\affiliation{Institute of Modern Physics, Chinese Academy of Sciences, Lanzhou 730000, China}

\author{Xingbo Zhao}
\affiliation{
  School of Nuclear Science and Technology, University of Chinese Academy of Sciences, Beijing 100049, China
}
\affiliation{Institute of Modern Physics, Chinese Academy of Sciences, Lanzhou 730000, China}
\affiliation{CAS Key Laboratory of High Precision Nuclear Spectroscopy, Institute of Modern Physics, Chinese Academy of Sciences, Lanzhou 730000, China}

\begin{abstract}
The heavy quarks present in the quark-gluon plasma (QGP) can act as a probe of relativistic heavy ion collisions as they retain the memory of their interaction history. In a previous study, a stochastic Schr\"{o}dinger equation (SSE) has been applied to describe the transport process of heavy quarks, where an external field with random phases is used to simulate the thermal medium. In this work, we study the connection between the SSE and the Boltzmann equation (BE) approach in the Keldysh Green's function formalism. By comparing the Green's function of the heavy quark from the SSE and the Keldysh Green's functions leading to the Boltzmann equation, we demonstrate that the SSE is consistent with the Boltzmann equation in the weak coupling limit. We subsequently confirm their consistency through numerical calculations.
\end{abstract}
\maketitle

\section{Introduction}
In ultra-relativistic heavy-ion collisions (URHICs), a significant portion of the kinetic energy 
from the colliding nuclei is transformed into thermal energy, leading to the creation of quark-gluon plasma (QGP)
from the strongly interacting matter. At high temperatures, quarks and gluons behave like free particles comprising the components of QGP. The expansion and cooling of QGP are driven by pressure, leading to the formation 
of the hadron gas. The thermalization time of QGP is typically within the range of 5-10 $\mathrm{fm}/c$ \cite{rapp2008heavy, yagi2005quark}.
While light quarks like \textit{u} and \textit{d} quarks thermalize at around 0.5 $\mathrm{fm}/ c$, which is much shorter than the lifetime of QGP, heavy quarks 
have a comparable lifetime due to their larger mass; thus, they retain the memory of their interaction history.
As a result, they act as a ``probe" of relativistic heavy ion collisions~\cite{zhao2011medium,du2017color,wu2021x,wu2023recombination}.
In recent years, various descriptions of heavy 
quarks' transport process in QGP (or other thermal media) have been developed, including the Boltzmann equation
(Fokker-Planck equation)~\cite{rapp2008heavy, he2013relativistic, yao2021coupled}, Schr\"{o}dinger-Langevin 
equation~\cite{kostin1972schrodinger, katz2016schrodinger}, and the Lindblad 
equation~\cite{akamatsu2018quantum, akamatsu2015heavy, de2017fate}.
Additionally, the Schwinger-Keldysh formalism has been used to describe the evolution of non-equilibrium 
systems \cite{schwinger1961brownian,keldysh1965diagram}. In this formalism, the Keldysh Green's functions are introduced. The Boltzmann equation can be derived by performing perturbative expansion of the Keldysh Green's functions. The Schwinger-Keldysh formalism serves as a bridge connecting quantum field theory and transport theory.

Although the Boltzmann equation can describe the evolution of heavy quarks in a thermal medium, it only takes into account a partial set of quantum effects through the scattering matrix element and the Pauli blocking factor. In order to overcome these limitations, the authors in~\cite{wupreparing} constructed a new approach to describe a heavy quark's evolution in a thermal medium using a stochastic Schr\"{o}dinger Equation (SSE). In that work, the authors introduced a classical gluon field to model the thermal medium and random phases for the gluon field to simulate the thermal fluctuation of the medium on the amplitude level. The interaction between this gluon field and the heavy quark was described using the SSE. On this basis, the authors in~\cite{wupreparing} further introduced a time correlation between random phases of the gluon field at different times, which characterizes the time scale for the evolution of the random phases of the gluon field. Based on numerically comparing the time evolution of the average momentum squared and average displacement squared with those from the Langevin equation approach, the authors suggested that the SSE approach can be used to describe the evolution of a heavy quark in the thermal medium. Because the SSE approach describes the evolution on the amplitude level, it can potentially capture more complete quantum effects compared to the Boltzmann equation (BE) approach. The explicit connection between the SSE and BE approach was however not shown in~\cite{wupreparing}.

In this paper, we demonstrate the connection between the SSE and the Boltzmann equation using the Schwinger-Keldysh formalism and subsequently verify their consistency through numerical calculations. We first briefly review the Schwinger-Keldysh approach in Sec.2. Then we demonstrate the connection between the Boltzmann equation and the stochastic Schr\"{o}dinger equation (SSE) in the Schwinger-Keldysh formalism in Sec.3. By considering the time correlation of random phases, we find that the SSE describes the process of an on-shell heavy quark absorbing (or emitting) an off-shell gluon and forming a final-state on-shell heavy quark. Next, we numerically calculate the evolution of the heavy quark momentum distributions in the Boltzmann equation and compare with those from the SSE in Sec.4. Finally, we conclude in Sec.5.

\section{The Keldysh Green's function and the Boltzmann equation}

\begin{figure}[!h]
  \includegraphics[scale=0.4]{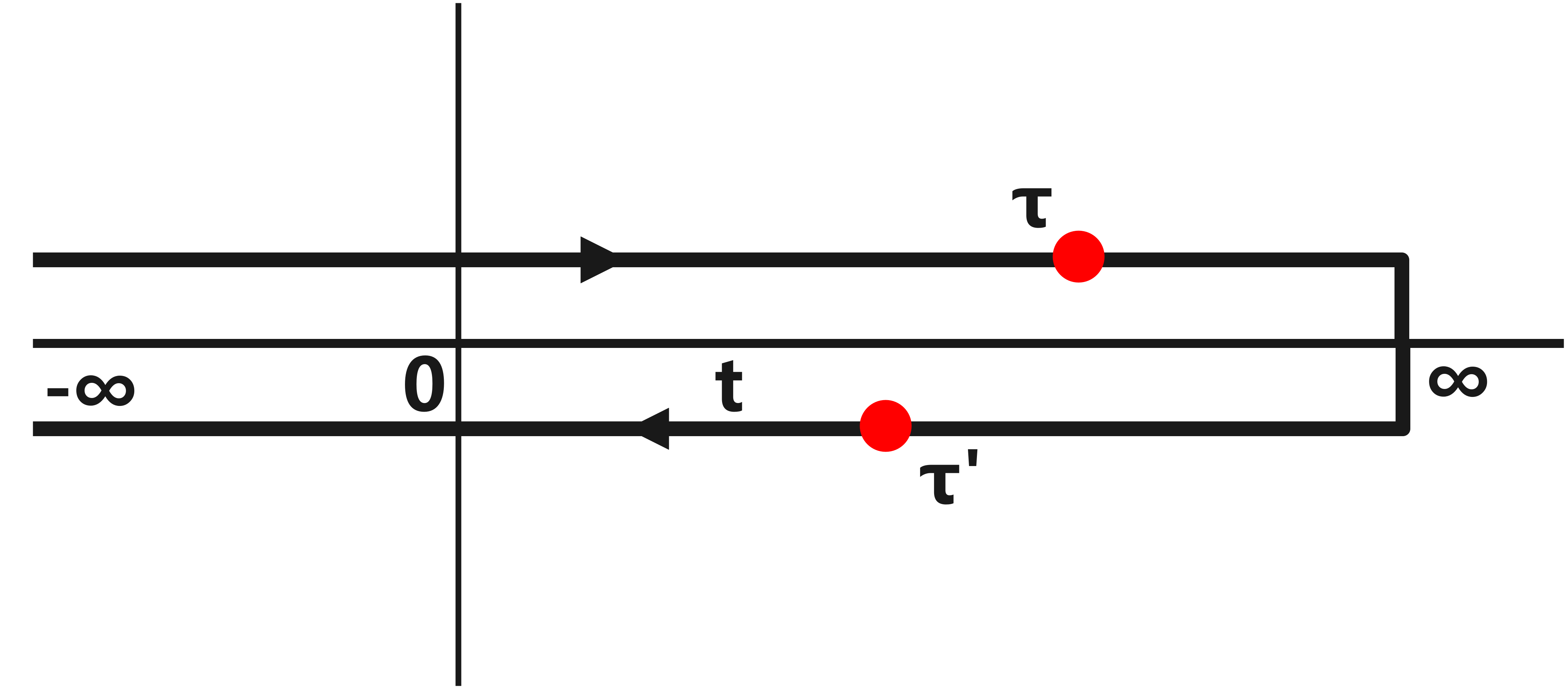} 
  \caption{The Keldysh time contour. $\tau$ and $\tau'$ denote the time coordinates of the Keldysh Green's functions, see text for the detail.}
  \label{time_integration_path}.
  \end{figure}

The Green's function (or propagator) is a mathematical tool in quantum field theory to calculate the probability of particle scattering and spectral distributions in various systems,
at both zero-temperature~\cite{peskin2018introduction} and finite temperature~\cite{kapusta1989finite}.
To extend the Green's functions to non-equilibrium systems, Schwinger and Keldysh introduced the Keldysh Green's functions, and the Keldysh time contour~\cite{schwinger1961brownian,keldysh1965diagram}, as illustrated in Fig.~(\ref{time_integration_path}). The Keldysh Green's functions play a central role in illustrating the connection between the SSE and the Boltzmann equation approach. Before introducing this connection, we briefly review the properties of the Keldysh Green's functions and the derivation of the Boltzmann equation using the Keldysh Green's functions~\cite{landau1981course,vspivcka2014electron}. 

The Keldysh Green's functions for spin-0 bosons are~\cite{landau1981course}:
 \begin{equation}
    \label{boson_keldysh_green_function}
  \begin{split}
  G^{--}_{B}(x,y)&=-i \langle \mathrm{T}[\phi(x)\phi^{\dagger}(y)]\rangle , \\
  G^{++}_{B}(x,y)&=-i \langle \tilde{\mathrm{T}}[\phi(x)\phi^{\dagger}(y)]\rangle ,\\
  G^{+-}_{B}(x,y)&=-i \langle \phi(x)\phi^{\dagger}(y)\rangle ,\\
  G^{-+}_{B}(x,y)&=-i \langle \phi^{\dagger}(y)\phi(x)\rangle .
  \end{split}
  \end{equation}
Here $x$ and $y$ denote the positions in space-time, $\langle...\rangle $ denotes the ensemble average and $T$ ($\tilde{T}$) denotes the (anti-)time-ordered product. For spin-$\frac{1}{2}$ fermions, the Green's functions are:
       \begin{equation}
    \label{keldysh_green_function}
  \begin{split}
  G^{--}_{F}(x,y)&=-i \langle \mathrm{T}[\psi(x)\bar{\psi}(y)]\rangle , \\
  G^{++}_{F}(x,y)&=-i \langle \tilde{\mathrm{T}}[\psi(x)\bar{\psi}(y)]\rangle ,\\
  G^{+-}_{F}(x,y)&=-i \langle \psi(x)\bar{\psi}(y)\rangle ,\\
  G^{-+}_{F}(x,y)&= i \langle \bar{\psi}(y)\psi(x)\rangle .
  \end{split}
  \end{equation}

In the above definitions of Green's functions, the boson field $\phi(x)=\phi(\vec{x},t)$ and its complex conjugation $\phi^{\dagger}(x)$ can be expanded in momentum space as
     \begin{equation}
     \begin{split}
        \phi(\vec{x},t)&=\int \frac{d^{3} \vec{p}}{(2\pi)^{3}\sqrt{2E_{\vec{p}}}} \left( \hat{a}_{\vec{p}} e^{i \vec{p} \cdot \vec{x}-iE_{\vec{p}}t}+\hat{a}^{\dagger}_{\vec{p}} e^{-i \vec{p} \cdot \vec{x}+iE_{\vec{p}}t} \right),\\
        \phi^{\dagger}(\vec{x},t)&=\int\frac{d^{3} \vec{p}}{(2\pi)^{3}\sqrt{2E_{\vec{p}}}} \left( \hat{a}^{\dagger}_{\vec{p}} e^{-i \vec{p} \cdot \vec{x}+iE_{\vec{p}}t}+\hat{a}_{\vec{p}} e^{i \vec{p} \cdot \vec{x}-iE_{\vec{p}}t} \right).
    \end{split}
    \label{boson_quantum_field}
  \end{equation}
Here $E_{\vec{p}}=\sqrt{\vec{p}^{2}+m^{2}}$ is the kinetic energy of the momentum mode $\vec{p}$, $m$ is the mass of the boson, $p=(E_{\vec{p}},\vec{p})$ is the 4-momentum of the
 mode $\vec{p}$. $\hat{a}_{\vec{p}}$ ($\hat{a}^{\dagger}_{\vec{p}}$) is the annihilation (creation) operator of the boson and satisfies the following commutation relation
\begin{equation}
    \left[\hat{a}_{\vec{p}}, \hat{a}^{\dagger}_{\vec{p}\,'}\right]=(2\pi)^{3} \delta^{3}(\vec{p}-\vec{p}\,').
\label{commutataion}
\end{equation}
The fermion fields $\psi(x)$ and $\bar{\psi}(x)$ can be similarly expanded as~\cite{peskin2018introduction}
     \begin{equation}
     \begin{split}
        \psi(\vec{x},t)&=\int \frac{d^{3} \vec{p}}{(2\pi)^{3}\sqrt{2E_{\vec{p}}}}\\
        &\,\sum\limits_{s}\left( \hat{b}^{s}_{\vec{p}} u^{s}(p) e^{i \vec{p} \cdot \vec{x}-iE_{\vec{p}}t}+\hat{d}^{s\dagger}_{\vec{p}} v^{s}(p) e^{-i \vec{p} \cdot \vec{x}+iE_{\vec{p}}t} \right),\\
        \bar{\psi}(\vec{x},t)&=\int\frac{d^{3} \vec{p}}{(2\pi)^{3}\sqrt{2E_{\vec{p}}}}\\
        &\,\sum\limits_{s}\left( \hat{b}^{s\dagger}_{\vec{p}} \bar{u}^{s}(p) e^{-i \vec{p} \cdot \vec{x}+iE_{\vec{p}}t}+\hat{d}^{s}_{\vec{p}} \bar{v}^{s}(p) e^{i \vec{p} \cdot \vec{x}-iE_{\vec{p}}t} \right).
    \end{split}
    \label{fermion_quantum_field}
  \end{equation}
Here $s$ is the spin index. $\hat{b}^{s}_{\vec{p}}$ ($\hat{b}^{s\dagger}_{\vec{p}}$) is the annihilation (creation) operator of the spin-$\frac{1}{2}$ fermion with momentum $\vec{p}$,  and $\hat{d}^{s}_{\vec{p}}$ ($\hat{d}^{s\dagger}_{\vec{p}}$) is the annihilation (creation) operator of the spin-$\frac{1}{2}$ antifermion. These operators satisfy the following anti-commutation relation
\begin{equation}
   \left\{\hat{b}^{r}_{\vec{p}},\hat{b}^{s\dagger}_{\vec{p}\,'}\right\}=\left\{\hat{d}^{r}_{\vec{p}},\hat{d}^{s\dagger}_{\vec{p}\,'}\right\}=(2\pi)^{3} \delta^{3}(\vec{p}-\vec{p}\,') \delta^{rs}.
\label{quark_commutataion}
\end{equation}
Here $u^{s}(p)$ and $ v^{s}(p), s=1,2$ are two linearly independent spinors for the fermion and antifermion, respectively. $u^{s}(p)e^{-ip\cdot x}$ and $v^s(p)e^{i p \cdot x}$ are the corresponding plane-wave solutions of the Dirac equation. 

In the definition of the Keldysh Green's functions in Eq.~(\ref{boson_keldysh_green_function}) and Eq.~(\ref{keldysh_green_function}), the superscripts "$+$" and "$-$" signify the respective branch of the time contour $\mathbb{C}$ where the time coordinates $\tau$ and $\tau'$  of the Green's function $G^{\alpha\beta}(\tau,\vec{x};\tau',\vec{x}\,')$ reside: ``$+$'' represents the branch from $\infty$ to $-\infty$ ($\mathbb{C}_{+}$), and ``$-$'' represents the branch from $-\infty$ to $\infty$ ($\mathbb{C}_{-}$), see Fig.~(\ref{time_integration_path}). For bosons, we have~\cite{kleinert2009path,van2014nonequilibrium,landau1981course}
  \begin{equation}
  \begin{split}
  \langle \mathrm{T}[\phi(x)\phi^{\dagger}(y)]\rangle =\left\{
  \begin{aligned}
   \langle \phi(x)\phi^{\dagger}(y)\rangle , \quad t_{x} >  t_{y} \\
   \langle \phi^{\dagger}(y)\phi(x)\rangle , \quad t_{x} <  t_{y},
  \end{aligned}
  \right.
  \end{split}
  \label{define_of_boson_T}
  \end{equation}
  where $t_{x}, t_{y}\in \mathbb{C}_{-}$, and
  \begin{equation}
  \begin{split}
  \langle \tilde{\mathrm{T}}[\phi(x)\phi^{\dagger}(y)]\rangle =\left\{
  \begin{aligned}
   \langle \phi^{\dagger}(y)\phi(x)\rangle , \quad t_{x} >  t_{y} \\
   \langle \phi(x)\phi^{\dagger}(y)\rangle , \quad t_{x} <  t_{y},
  \end{aligned}
  \right.
  \end{split}
  \label{define_of_boson_tildeT}
  \end{equation}
where $t_{x}, t_{y}\in \mathbb{C}_{+}$. For fermions, we have
\begin{equation}
    \begin{split}
  \langle \mathrm{T}[\psi(x)\bar{\psi}(y)]\rangle =\left\{
  \begin{aligned}
   \langle \psi(x)\bar{\psi}(y)\rangle , \quad t_{x} >  t_{y} \\
   -\langle \bar{\psi}(y)\psi(x)\rangle , \quad t_{x} < t_{y},
  \end{aligned}
  \right.
  \end{split}
  \label{define_of_fermion_T}
\end{equation}
\begin{equation}
    \begin{split}
  \langle \tilde{\mathrm{T}}[\psi(x)\bar{\psi}(y)]\rangle =\left\{
  \begin{aligned}
   -\langle \bar{\psi}(y)\psi(x)\rangle , \quad t_{x} >  t_{y} \\
   \langle \psi(x)\bar{\psi}(y)\rangle , \quad t_{x} < t_{y}.
  \end{aligned}
  \right.
  \end{split}
  \label{define_of_fermion_tildeT}
\end{equation}

By substituting the expansions of $\phi$ and $\phi^{\dagger}$ into the definition of Green's functions, we can obtain the non-interacting boson Green's functions $G^{-+}_{B}$ and $ G^{+-}_{B}$~\cite{van2014nonequilibrium}:
  \begin{equation}
  \label{G-+}
  \begin{split}
  &G_{B}^{-+}(x,y)= -2\pi i \\
  &\times \int \frac{d^{3}\vec{p} dE}{(2\pi)^4} \big\{ n_{B}(\vec{p}) e^{i\vec{p} \cdot \Delta \vec{x} -i E \Delta t} \theta(E)\delta(E^2-\vec{p}^2-m^2) \\
  &\quad + \left[n_{B}(\vec{p})+1\right] e^{-i\vec{p} \cdot \Delta \vec{x} +i E \Delta t}\theta(E)\delta(E^2-\vec{p}^2-m^2) \big\}, \\
  &G_{B}^{+-}(x,y)= -2\pi i \\
  &\times \int \frac{d^{3}\vec{p} dE}{(2\pi)^4} \big\{  n_{B}(\vec{p}) e^{i E \Delta t-i\vec{p} \cdot \Delta \vec{x} }\theta(E)\delta(E^2-\vec{p}^2-m^2) \\
  &\quad +\left[n_{B}(\vec{p})+1\right] e^{-i E \Delta t+i\vec{p} \cdot \Delta \vec{x} } \theta(E)\delta(E^2-\vec{p}^2-m^2) \big\}.
  \end{split}
  \end{equation}
In Eq.~(\ref{G-+}), $n_{B}(\vec{p})=\bra{\Omega}\hat{a}^{\dagger}_{\vec{p}}\hat{a}_{\vec{p}}\ket{\Omega}$ is the momentum distribution of the bosons in the system $\ket{\Omega}$. The term $n_{B}(\vec{p})+1$ follows from the commutation relation in Eq.~(\ref{commutataion}). Here $x=(t_{x},\vec{x})$ and $y=(t_{y},\vec{y})$ denote the positions in 4-dimensional space-time. $\Delta \vec{x}=\vec{x}-\vec{y}$ and $\Delta t=t_{x}-t_{y}$ are their differences.

The non-interacting fermion Keldysh Green's functions can be similarly derived by expanding the quark field. Since in~\cite{wupreparing} only a single fermion is considered in the SSE, in this study we accordingly neglect the contribution from the antifermions. According to the definitions of the fermion Keldysh Green's function, we then have
\begin{equation}
\begin{split}
    &G_{F,\xi\xi'}^{-+}(x,y)=2\pi i \left(i\gamma \cdot \partial_{x} +m \right)_{\xi\xi'} G_{F,0}^{-+}(x,y)\\
    &=  2\pi i \left(i\gamma \cdot \partial_{x} +m \right)_{\xi\xi'} \\
    &\times \int \frac{d^{3}\vec{p} dE}{(2\pi)^4} n_{F}(\vec{p}) e^{i\vec{p} \cdot \Delta \vec{x} -i E \Delta t} \theta(E)\delta(E^2-\vec{p}^2-m^2), \\
    &G_{F,\xi\xi'}^{+-}(x,y)=-2\pi i \left(i\gamma \cdot \partial_{x} +m \right)_{\xi\xi'} G_{F,0}^{+-}(x,y)\\
    &=  -2\pi i \left(i\gamma \cdot \partial_{x} +m \right)_{\xi\xi'} \\
    &\times \int \frac{d^{3}\vec{p} dE}{(2\pi)^4} \big\{\left[1-n_{F}(\vec{p}) \right] e^{i\vec{p} \cdot \Delta \vec{x} -i E \Delta t}\\ 
&\times \theta(E)\delta(E^2-\vec{p}^2-m^2)\big\}.
\end{split}
\label{fermion_keldysh_green}
\end{equation}
Here $G^{-+}_{F,0}$ and $G^{+-}_{F,0}$ denote the fermion Keldysh Green's function without the spinor part. $n_{F}(\vec{p})=\sum\limits_{s=1,2}\bra{\Omega}\hat{b}^{s+}_{\vec{p}}\hat{b}^{s}_{\vec{p}}\ket{\Omega}$ is the momentum distribution of the fermions in the system $\ket{\Omega}$. $\xi$ and $\xi'$ are the spinor indices. $\gamma$ in Eq.~(\ref{fermion_keldysh_green}) represents the Dirac matrices $\gamma^{\mu}$, and $\mu=0,1,2,3$. $\partial_{x}=\partial_{x,\mu}$ denotes the derivative with respect to the space-time coordinate $x$, and $m$ is the mass of the fermion. 

\begin{figure}
     \centering
  \begin{tikzpicture}
  \begin{feynman}
\vertex (a1);
\vertex[right=1.5cm of a1] (b1);
\vertex[right=3.5cm of a1] (a2);
\vertex[right=1.5cm of a2] (b2);
\vertex[below=2.0cm of a1] (a3);
\vertex[right=2.0cm of a3] (b3);
\vertex[below=1.5cm of a2] (c1);
\vertex[below=1.0cm of c1] (c2);
\vertex[below=2.0cm of a3] (a4);
\vertex[below=2.0cm of b3] (b4);
\vertex[below=2.0cm of c1] (c3);
\vertex[below=2.0cm of c2] (c4);
 \diagram* {
  (a1) -- [fermion, momentum=\(p\)] (b1),
  };
  \node at (0.0,-0.5) {$\beta$};
  \node at (1.5,-0.5) {$\alpha$};
  \node at (2.3,0.0) {$=iG_{F}^{\alpha \beta}(p)$};
\diagram* {
  (a2) -- [boson, momentum=\(p\)] (b2),
  };
   \node at (3.5,-0.5) {$\beta$};
  \node at (5.0,-0.5) {$\alpha$};
  \node at (5.8,0.0) {$=iG_{B}^{\alpha \beta}(p)$};
\diagram* {
  (a3) -- [boson, momentum=\(p\)] (b3),
  {(b3) -- [fermion, momentum=\(k'\)] (c1)},
  {(c2) -- [fermion, momentum=\(k\)] (b3)},
  };
  \node at (1.5,-2.5) {$+$};
  \node at (5.0,-2.0) {$=ig\gamma^{\mu} \delta^{4}(k+p-k')$};
\diagram* {
  (a4) -- [boson, momentum=\(p\)] (b4),
  {(b4) -- [fermion, momentum=\(k'\)] (c3)},
  {(c4) -- [fermion, momentum=\(k\)] (b4)},
  };
  \node at (1.5,-4.5) {$-$};
  \node at (5.0,-4.0) {$=-ig\gamma^{\mu} \delta^{4}(k+p-k')$};
  \end{feynman}
  \end{tikzpicture}
    \caption{Feynman rules for the Keldysh Green's functions~\cite{landau1981course}.}
    \label{fig:element_feynman_physical_correspondence}
\end{figure}
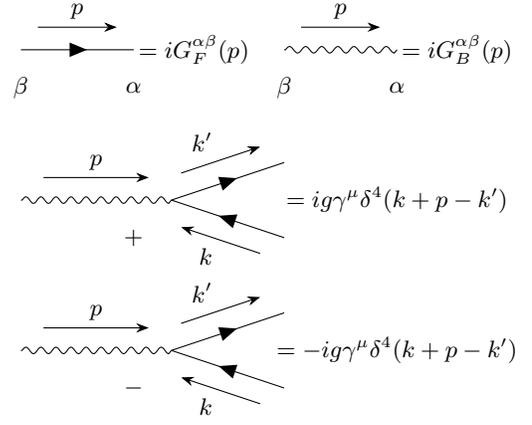

The interacting Keldysh Green's functions can be expanded in perturbation theory in terms of the Feynman diagrams. In Fig.~(\ref{fig:element_feynman_physical_correspondence}), we provide the Feynman rules, which are adopted from~\cite{landau1981course}: the solid line represents the fermions, and the wavy line represents the bosons. $\alpha,\beta=\pm$ at the end and start points of the lines are the labels for the doublet fermion or boson field at finite temperature. At the vertices, the signs of the labels can be either $+$ or $-$. $g$ is the coupling constant between the fermion and boson field, see Sec. 3 for the details.

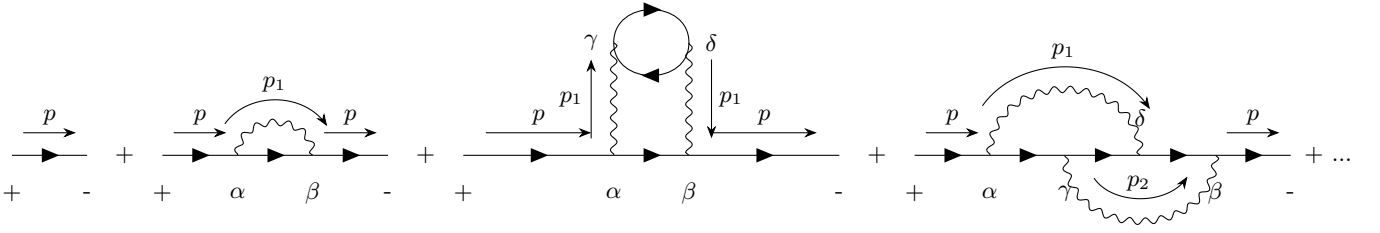
\begin{figure*}[t]
  \centering
  \begin{tikzpicture}
  \begin{feynman}
  \vertex (a1);
  \vertex[right=2.0cm of a1](a2);
  \vertex[right=6.0cm of a1](a3);
  \vertex[right=12.0cm of a1](a4);
  \vertex[right=1.0cm of a1] (b1);
    \vertex[right=1.0cm of a2] (b2);
      \vertex[right=2.0cm of a3] (b3);
        \vertex[right=1.0cm of a4] (b4);
  \vertex[right=1.0cm of b2] (c2);
    \vertex[right=1.0cm of b3] (c3);
      \vertex[right=1.0cm of b4] (c4);
  \vertex[right=1.0cm of c2] (d2);    
    \vertex[right=2.0cm of c3] (d3);
      \vertex[right=1.0cm of c4] (d4);
  \vertex[right=1.0cm of d4] (e4);
  \vertex[right=1.0cm of e4] (f4);
  \vertex[above=1.5cm of b3] (z3);
  \vertex[above=1.5cm of c3] (y3);
  \diagram* {
  (a1) -- [fermion, momentum=\(p\)] (b1),
  };
  \node at (0.0,-0.5) {+};
  \node at (1.0,-0.5) {-};
  \node at (1.5,0) {+};
  \diagram* {
  (a2) -- [fermion, momentum=\(p\)] (b2) -- [fermion] (c2) -- [fermion, momentum=\(p\)] (d2),
  {(b2) -- [boson, half left, momentum=\(p_{1}\)] (c2)},
  };
  \node at (2.0,-0.5) {+};
  \node at (3.0,-0.5) {$\alpha$};
  \node at (4.0,-0.5) {$\beta$};
  \node at (5.0,-0.5) {-};
  \node at (5.5,0) {+};
  \diagram* {
  (a3) -- [fermion, momentum=\(p\)] (b3) -- [fermion] (c3) -- [fermion, momentum=\(p\)] (d3),
  {(b3) -- [boson, momentum=\(p_{1}\)] (z3)},
  {(y3) -- [boson, momentum=\(p_{1}\)] (c3)},
  (z3) -- [fermion, half left] (y3) -- [fermion, half left] (z3),
  };
  \node at (6.0,-0.5) {+};
  \node at (8.0,-0.5) {$\alpha$};
  \node at (7.7,1.5) {$\gamma$};
  \node at (9.0,-0.5) {$\beta$};
  \node at (9.3,1.5) {$\delta$};
  \node at (11.0,-0.5) {-};
  \node at (11.5,0) {+};
  \diagram* {
  (a4) -- [fermion, momentum=\(p\)] (b4) -- [fermion] (c4) -- [fermion] (d4) -- [fermion] (e4) -- [fermion, momentum=\(p\)] (f4),
  {(b4) -- [boson, half left, momentum=\(p_{1}\)] (d4)},
  {(c4) -- [boson, half right, momentum=\(p_{2}\)] (e4)},
  };
  \node at (12.0,-0.5) {+};
  \node at (13.0,-0.5) {$\alpha$};
  \node at (14.0,-0.5) {$\gamma$};
  \node at (15.0,0.5) {$\delta$};
  \node at (16.0,-0.5) {$\beta$};
  \node at (17.0,-0.5) {-};
  \node at (17.5,0) {+ ...};
  \end{feynman}
  \end{tikzpicture}
\caption{The perturbative expansion of the interacting Keldysh Green's function $G^{-+}_{F}$ in a medium. The solid lines represent the noninteracting Keldysh Green's functions for the fermion. The wavy lines represent the noninteracting Keldysh Green's functions for the boson. $\alpha, \beta, \gamma, \delta = +, -$ are the labels of different kinds of Keldysh Green's functions in Eq.~(\ref{keldysh_green_function}).}
  \label{fig_green_function}
\end{figure*}

\begin{figure}[!h]
  \centering
  \begin{tikzpicture}
  \begin{feynman}
  \vertex (a1);
  \vertex[above=1.0cm of a1] (b1);
  \vertex[above=1.0cm of b1] (c1);
  \vertex[left=1.0cm of a1] (d1);
  \vertex[right=1.0cm of a1] (e1);
  \diagram* {
  (a1) -- [boson] (b1),
  (b1) -- [plain, half left] (c1),
  (c1) -- [fermion, half left] (b1),
  (d1) -- [fermion,momentum=\(p\)] (a1),
  (a1) -- [fermion,momentum=\(p\)] (e1),
  };
  \node at (-1,-0.5) {$+$};
  \node at (1,-0.5) {$-$};
  \node at (0,-0.5) {$\alpha$};
  \node at (-0.2, 0.8) {$\beta$};
  \end{feynman}
  \end{tikzpicture}
  \caption{The tadpole contribution to the Keldysh Green's function $G^{-+}_{F}$. This contribution vanishes in a symmetric medium with respect to fermions and antifermions. The solid line represents the Green's function for the fermion, and the wavy line represents the Green's function for the boson.}
  \label{excluded_term}
\end{figure}
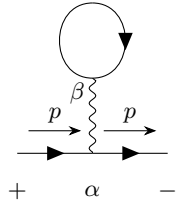

For the fermion field, the perturbative expansion of the interacting two-point Keldysh Green's functions is illustrated in Fig.~(\ref{fig_green_function})~\cite{landau1981course, geiger1996quantum}. We ignore the term represented in Fig.~(\ref{excluded_term}) because in this work we assume the thermal medium to be symmetric with respect to particles and antiparticles for simplicity.

Now we are ready to derive the Boltzmann equation for the fermions using the Keldysh Green's functions. The following derivation follows from~\cite{landau1981course}. We repeat it here for the convenience of the readers.

\begin{figure}[!h]
  \centering
  \begin{tikzpicture}
  \begin{feynman}
  \vertex (a1);
  \vertex[right=1.0cm of a1] (b1);
  \vertex[right=1.0cm of b1] (c1);
  \vertex[right=1.0cm of c1] (d1);
  \vertex[right=1.0cm of d1] (e1);
  \vertex[right=1.0cm of e1] (f1);
  \diagram* {
  (a1) -- [fermion] (b1),
  (a1) -- [boson, half left] (b1),
  };
  \node at (0,-0.5) {$\beta$};
  \node at (1,-0.5) {$\alpha$};
  \node at (1.5,0) {$+$};
  \diagram* {
  (c1) -- [fermion] (d1),
  (d1) -- [fermion] (e1),
  (e1) -- [fermion] (f1),
  (c1) -- [boson, half left] (e1),
  (d1) -- [boson, half left] (f1),
  };
  \node at (2.0,-0.5) {$\beta$};
  \node at (3.0,-0.5) {$\gamma$};
  \node at (4.0,-0.5) {$\delta$};
  \node at (5.0,-0.5) {$\alpha$};
  \node at (5.5,0) {$+$};
  \node at (6.0,0) {...};
  \end{feynman}
  \end{tikzpicture}
  \caption{Feynman diagrams of the one-particle-irreducible diagrams of the self-energy $\Sigma^{\alpha \beta}$ for the fermion.}
  \label{irreducible_parts}
  \end{figure}
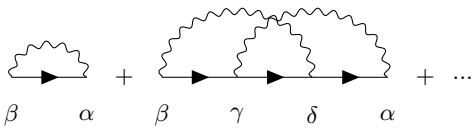

To obtain the Boltzmann equation, we first apply the differential operator $i\gamma\cdot \partial_{x}-m$ from the Dirac equation to the fermion Keldysh Green's function $G_{F}^{-+}$, and obtain its equation of motion in 4-dimensional space-time. To simplify the derivation, we introduce the differential operator $G^{-1}_{x,\xi\xi'}=-\left(\partial^{\mu}_{x}\partial_{x,\mu}+m^{2}\right) \mathbf{I}_{\xi\xi'}$ ($\mathbf{I}$ is a $4\times 4$ identity matrix in the spinor space), which is the product of $i\gamma\cdot \partial_{x}-m$ and the operator $i\gamma\cdot \partial_{x}+m$. Applying the operator $G^{-1}_{x,\xi\xi'}$ to the Keldysh Green's function $G^{-+}_{F,0}$ and making use of the equation of motion for the fermion operator $\psi(x)$, we obtain the following equation of motion ~\cite{landau1981course}
\begin{equation}
\begin{split}
   &G^{-1}_{x,\xi\xi'} G_{F,0}^{-+}(x,y) \\
   &=\sum\limits_{\alpha=+,-} \sum\limits_{\xi''}\int d^{4}x' \Sigma^{-\alpha}_{F,\xi\xi''}(x,x')G_{F,\xi''\xi'}^{\alpha+}(x',y).
   \end{split}
\label{motion_equation_G}
\end{equation}
Here $x, y, x'$ are the space-time coordinates. $\Sigma^{-\alpha}_{F,\xi\xi''}$ is the fermion self-energy multiplied by the imaginary unit $-i$, which is necessary for the compatibility between the definition of the self-energy and its Feynman diagram representation. The fermion's self-energy $\Sigma^{\alpha\beta}$ consists of the sum of one-particle-irreducible diagrams, see Fig.~(\ref{irreducible_parts})~\cite{geiger1996quantum}. For convenience, we omit the spinor indices $\xi$ and $\xi'$ in the following part of this paper.

Now, we perform the Wigner transformation on both sides of the following equation, which follows from Eq.~(\ref{motion_equation_G})~\cite{landau1981course},
\begin{equation}
\begin{split}
&(G^{-1}_{y}-G^{-1}_{x})G_{F,0}^{-+}(x,y) \\
&=-\sum\limits_{\alpha=+,-}\int d^{4}x' \big[\Sigma^{-\alpha}(x,x')G_{F}^{\alpha+}(x',y) \\
&\quad +G_{F}^{-\alpha}(x,x')\Sigma^{\alpha+}(x',y) \big].
\end{split}
\label{G_G}
\end{equation}
In this paper, we adopt the following definition for the Wigner transformation in 4-dimensional space-time for the fermion Green's functions \cite{landau1981course,gamble2008foundations}
\begin{equation}
\begin{split}
    &\mathcal{W}\left[ G_{F}^{-+}(x,y) \right] \\
    &=\mathcal{W}\left[ G_{F}^{-+}(X+\frac{1}{2}x',X-\frac{1}{2}x') \right] \\
    &=\int \frac{d^{4}x'}{(2\pi)^{4}} e^{iP\cdot x'} G_{F}^{-+}(X+\frac{1}{2}x',X-\frac{1}{2}x').
\end{split}
\label{wigner_define}
\end{equation}
Here $\mathcal{W}$ denotes the Wigner transformation and $X=\frac{(x+y)}{2}, x'=x-y$. $P=(P^{0},\vec{P})$ is the 4-momentum of the fermion. Next we will show that after the Wigner transformation, the left-hand side of Eq.~(\ref{G_G}) is identified as the drift term of the Boltzmann equation, and the right-hand side becomes the scattering term. 

If we assume that the fermion Green's function $G_{F}^{-+}(X+\frac{x'}{2},X-\frac{x'}{2})$ weakly depends on the position $X$~\cite{landau1981course}, we have
\begin{equation}
\begin{split}
    &\int \frac{d^{4}x'}{(2\pi)^{4}} e^{iP\cdot x'} G_{F}^{-+}(X+\frac{1}{2}x',X-\frac{1}{2}x') \\
    &\approx \int \frac{d^{4}x'}{(2\pi)^{4}} e^{iP\cdot x'} G_{F}^{-+}(\frac{1}{2}x',-\frac{1}{2}x') \\
    &=\int \frac{d^{4}x'}{(2\pi)^{4}} \int \frac{d^{3}\vec{p}dE}{(2\pi)^{4}}
        (i\gamma\cdot\partial_{x'}+m)n_{F}(\vec{p}) e^{iP\cdot x'} e^{i\vec{p}\cdot\vec{x}'-iEt'} \\
    &\quad \times(2\pi i)\theta(E)\delta(E^{2}-\vec{p}^{2}-m^{2}) \\
    &=\frac{1}{(2\pi)^{4}} \int \frac{d^{3}\vec{p}dE}{(2\pi)^{4}} (i\gamma\cdot p+m)n_{F}(\vec{p}) \\
    &\quad \times (2\pi)^{5} i\theta(E)\delta(E^{2}-\vec{p}^{2}-m^{2}) \delta^{3}(\vec{P}-\vec{p})\delta(P^{0}-E) \\
    &=\frac{i(i\gamma\cdot P+m)}{(2\pi)^{3}} n_{F}(\vec{P})\theta(P^{0})\delta\left[\left(P^{0}\right)^{2}-\vec{P}^{2}-m^{2}\right].
\end{split}
\label{wigner_Green's_function}
\end{equation}
Here $p=(E,\vec{p})$ is the 4-momentum in the fermion Green's function. In the second equality, we have used the Fourier expansion of the fermion Green's function in Eq.~(\ref{fermion_keldysh_green}). From this derivation, we find that the Wigner transformation of $G^{-+}_{F}(x,y)$ is approximately equal to the 4-dimensional Fourier transformation of $G^{-+}_{F}(\frac{1}{2}x',-\frac{1}{2}x')$ multiplied by $(2\pi)^{-4}$, where the weak dependence on $X$ is carried by $n_F(\vec{P})$. For the fermion systems weakly depending on the position $X$, its phase space distribution $f(X,P)$ approximately satisfies
\begin{equation}
f(X,P)\approx n_{F}(\vec{P})2E_{\vec{P}}\theta(P^{0})\delta\left[\left(P^{0}\right)^{2}-\vec{P}^{2}-m^{2}\right].
\label{approximation}
\end{equation}
Here $n_F(\vec{P})$ implicitly carries a weak dependence on the position $X$ of the fermion. Comparing Eq. (\ref{wigner_Green's_function}) and Eq. (\ref{approximation}), we identify the relation between the fermion's phase space distribution and the Wigner transformation of $G_F^{-+}$ as
\begin{equation}
    \mathcal{W}\left[ G_{F}^{-+}(x,y) \right]=\frac{i(i\gamma\cdot P+m)}{(2\pi)^{3}2E_{\vec{P}}} f(X,P).
\end{equation}
The Wigner transformation of the Green's function without the spinor part $G^{-+}_{F,0}$ is accordingly
\begin{equation}
\begin{split}
    &\mathcal{W}\left[G^{-+}_{F,0}(x,y)\right]\\
    &\approx \frac{i}{(2\pi)^{3}} n_{F}(\vec{P})\theta(P^{0})\delta\left[\left(P^{0}\right)^{2}-\vec{P}^{2}-m^{2}\right]\\
    &\approx \frac{i}{(2\pi)^{3}2E_{\vec{P}}} f(X,P).
\end{split}
\label{Wigner_G_F0}
\end{equation}

Similarly, the Wigner transformation of $G_{F,0}^{+-}$ is
\begin{align}
    \mathcal{W}\left[G^{+-}_{F,0}(x,y)\right]\approx \frac{i}{(2\pi)^{3}2E_{\vec{P}}} \left[1-f(X,P)\right].
\label{Wigner_G+-_F0}
\end{align}

Now, we are ready to perform the Wigner transformation on the left-hand side of Eq.~(\ref{G_G}). In position space, any two-variable function $\rho(x^{i},y^{i})$ satisfies the following relation~\cite{gamble2008foundations,cercignani2002relativistic,liboff2003kinetic} for the Wigner transformation (where $i=1,2,3$ is the label for the spatial directions):
\begin{equation}
\begin{split}
    \mathcal{W} \left[i \left(\partial^{2}_{x,i}-\partial_{y,i}^{2}\right)\rho(x^{i},y^{i})\right]&=-2P^{i} \frac{\partial}{\partial X^{i}} S(X^{i},P^{i}), \\
    S(X^{i},P^{i})&=\mathcal{W}\left[\rho(x^{i},y^{i})\right].
\end{split}
\label{property_of_wigner_transform}
\end{equation}
Here $\mathcal{W}$ denotes the Wigner transformation and $S(X^{i},P^{i})$ is the Wigner transformation of $\rho(x^{i},y^{i})$ and $X^i=\frac{(x^{i}+y^{i})}{2}$. $x^{i}$ and $ y^{i}$ are the $i$th spatial components of the 4-position $x$ and $y$, respectively. $P^{i}$ is the $i$th spatial component of the 4-momentum. For the time component of $x$ and $y$, the Wigner transformation is
\begin{equation}
\begin{split}
    &\mathcal{W}\left[i\left(\partial^{2}_{x,0}-\partial^{2}_{y,0}\right)\rho(x^{0},y^{0})\right]\\
    &=\int \frac{d(t_{x}-t_{y})}{2\pi} e^{iP^{0}(t_{x}-t_{y})} \left[i\left(\partial^{2}_{x,0}-\partial^{2}_{y,0}\right)\rho(x^{0},y^{0})\right]\\
    &=2P^{0}\frac{\partial}{\partial X^{0}} S(X^{0},P^{0}).
\end{split}
\label{time_wigner}
\end{equation}
Here we have used the relation $e^{i P \cdot x'}=e^{iP^{0}t'-i\vec{P}\cdot \vec{x}\,'}$ for the 4-momentum $P$ and 4-position $x'$.

By using Eq.~(\ref{property_of_wigner_transform}) and Eq.~(\ref{time_wigner}), the Wigner transformation on the left-hand side of Eq.~(\ref{G_G}) is
  \begin{equation}
  \begin{split}
  &\mathcal{W}\left[ \left(G^{-1}_{y}-G^{-1}_{x}\right)G_{F,0}^{-+}(x,y)\right]\\
  &=\mathcal{W}\left[\mathbf{I} \left[ \left(\partial_{x}^{\mu}\partial_{x,\mu}+m^{2}\right)-\left(\partial_{y}^{\mu}\partial_{y,\mu}+m^{2}\right)\right] G_{F,0}^{-+}(x,y) \right] \\
  &=2\left(P^{0}\frac{\partial}{\partial X^{0}}+\sum\limits_{j=1}^{3}P^{j}\frac{\partial}{\partial X^{j}}\right)\mathcal{W}\left[-iG_{F,0}^{-+}(x,y)\right]\mathbf{I},
  \label{left_hand_boltzmann}
  \end{split}
  \end{equation}
where $\bf{I}$ is the 4$\times$4 identity matrix in the spinor space. After integrating $P^{0}$, the result is
\begin{equation}
  \mathcal{W}\left[ \left(G^{-1}_{y}-G^{-1}_{x}\right)G_{F,0}^{-+}(x,y)\right]
  =\frac{\mathbf{I}}{(2\pi)^{3}E_{\vec{P}}} P^{\nu}\frac{\partial f(X,P)}{\partial X^{\nu}}.
  \label{simply_left_hand_boltzmann}
  \end{equation}
Here we have used the result of the Wigner transformation of $G_{F}^{-+}$ in Eq.~(\ref{Wigner_G_F0}).

  Next, we perform a Wigner transformation on the right-hand side of Eq.~(\ref{G_G}). We first simplify the right-hand side using the following relationship between the Keldysh Green's functions and fermion self-energies~\cite{landau1981course},
\begin{equation}
\begin{split}
&G^{--}_{F}(x,y)+G^{++}_{F}(x,y)\\
&\quad -G^{-+}_{F}(x,y)-G^{+-}_{F}(x,y)=0,\\
&\Sigma^{--}(x,y)+\Sigma^{++}(x,y)\\
&\quad +\Sigma^{-+}(x,y)+\Sigma^{+-}(x,y)=0.
\end{split}
\label{identity_G}
\end{equation}

As shown above, the Wigner transformation of a function is approximately equal to its Fourier transformation up to the integration over 4-position and a constant factor of $(2\pi)^{-4}$. Therefore, the Wigner transform of the convolution of the two functions is approximately equal to $(2\pi)^{4}$ times the product of each Wigner transformation. 

Finally, after substituting Eq.~(\ref{identity_G}) into the right side of Eq.~(\ref{G_G}), we obtain the Wigner transformation result of Eq.~(\ref{G_G}) as~\cite{landau1981course}
\begin{equation}
\begin{split}
\mathbf{I}P^{\mu}\frac{\partial f(X,P)}{\partial X^{\mu}} &= (2\pi)^{3} E_{\vec{P}}(2\pi)^{4} \\
&\times \big[-\Sigma^{-+}_{w}(X,P)G^{+-}_{F,w}(X,P) \\
&\quad +\Sigma^{+-}_{w}(X,P)G^{-+}_{F,w}(X,P)\big].
\end{split}
\label{G_G_simplify}
\end{equation}
Here $G_{F,w}^{+-}(X, P)$, $G_{F,w}^{-+}(X, P)$, $\Sigma_{w}^{+-}(X, P)$, $\Sigma_{w}^{-+}(X, P)$ are the Wigner transformation of $G^{+-}_{F}(x, y)$, $G^{-+}_{F}(x, y)$, $\Sigma^{+-}(x, y)$, $\Sigma^{-+}(x, y)$ respectively.

By comparing with the standard form of the relativistic Boltzmann equation~\cite{van2015introduction}
\begin{equation}
\begin{split}
    P^{\mu}\frac{\partial f(X,P)}{\partial X^{\mu}}&=\frac{1}{2(2\pi)^{3}} \int \frac{d^{3}\vec{P}_{1}}{E_{\vec{P}_{1}}}\int \frac{d^{3}\vec{P}\,'}{E_{\vec{P}\,'}}\int\frac{d^{3}\vec{P}_{1}\,'}{E_{\vec{P}\,'_{1}}} \\
    & \quad \times W(P',P'_{1}, P,P_{1})\left(f'f'_{1}-ff_{1}\right),
\label{relativistic_Boltzmann_van_Hees}
\end{split}
\end{equation}
we identify the right-hand side of Eq.~(\ref{G_G_simplify}) as the scattering term in the absence of the 4-force. In Eq.~(\ref{relativistic_Boltzmann_van_Hees}), $f,f_{1},f',f'_{1}$ are the phase space distribution functions of the incoming and outgoing particles in the scattering process at the position $X$ with momentum $\vec{P},\vec{P}_{1},\vec{P}\,',\vec{P}\,'_{1}$ respectively. $E_{\vec{P}\,'}, E_{\vec{P}_{1}},E_{\vec{P}_{1}\,'}$ are the corresponding energy of the particles participating in the scattering. The function $W$ is the scattering probability, which is encoded in the self-energies $\Sigma^{+-}_{w}$ and $\Sigma^{-+}_{w}$ in Eq.~(\ref{G_G_simplify})~\cite{landau1981course,vspivcka2014electron,geiger1996quantum}.

The exact relation between the right-hand side of Eq.~(\ref{G_G_simplify}) and (\ref{relativistic_Boltzmann_van_Hees}) will be investigated in the next section.

\section{Stochastic Schr\"{o}dinger equation}

In reference~\cite{wupreparing}, the authors introduced a stochastic Schr\"{o}dinger equation to describe the time evolution of a heavy quark in a thermal medium, which takes the following form:
\begin{equation}
  \label{original_sto_sch_equ}
  \begin{split}
  i \frac{\partial}{\partial t} \varphi(\vec{x},t)&=(H_{0}+H_{I}) \varphi(\vec{x},t), \\
  H_{0}&=\sqrt{-\nabla^{2}+m_{q}}, \\
  H_{I}&=\int d^{3}\vec{x} \,g\bar{\psi}\gamma^{\mu}\psi A^{\mu}.
  \end{split}
  \end{equation}
Here $\varphi(\vec{x},t)$ is the wave function of the heavy quark, $H_0=\sqrt{-\nabla^{2}+m_{q}^{2}}$ is the heavy quark's kinetic energy, $m_q$ is the mass of the heavy quark, and $H_{I}$ describes the interaction between the heavy quark and the thermal medium with the color degrees of freedom being ignored for simplicity. To account for the interaction between the heavy quark and the thermal medium, the authors modeled the medium as an external gluon field $A^{\mu}(\vec{x},t)$. In $H_{I}$, $g$ is the coupling constant between the quark field and the gluon field. Since the mass of the heavy quark is much larger than the typical temperature of the medium, the authors employed a non-relativistic approximation and considered only the coupling between the heavy quark and the zeroth component of the background field $A^{0}(\vec{x},t)$~\cite{wupreparing,peskin2018introduction}. For conciseness, we denote $A^{0}$ as $A$ in the following part of this paper. By expanding this field on the plane-wave basis, the authors obtained
      \begin{equation}
  \label{A field_expand}
  \begin{split}
  &A(\vec{x},t)=\int \frac{d^{3}\vec{p}}{(2\pi)^{3}} A(\vec{p},t) e^{i\vec{p}\cdot \vec{x}}=\int \frac{d^{3}\vec{p}}{(2\pi)^{3}} \sqrt{\frac{2d_{g}}{E_{\vec{p}}}}a(\vec{p},t) e^{i\vec{p}\cdot \vec{x}},\\
  &a(\vec{p},t)=\sqrt{n(\vec{p})} e^{i \theta(\vec{p},t)},\\
  &n(\vec{p})=\exp{\left(-\frac{E_{\vec{p}}}{T}\right)}=\exp{\left(-\frac{\sqrt{\vec{p}^{2}+m_{g}^2}}{T}\right)}.
  \end{split}
  \end{equation}
 Here $A(\vec{p},t)$ denotes the $A$ field in momentum space. The gluon momentum distribution $n(\vec{p})$ is assumed to (approximately) follow the Boltzmann distribution. $E_{\vec{p}}$ represents the on-shell gluon kinetic energy. The gluon thermal mass is denoted as $m_{g}=g_{0}\sqrt{1+\frac{N_{f}}{6}}T$. In this work we take $g_{0}$ as a strong coupling constant independent of the coupling constant  describing the interaction between the heavy quark and the medium, that is, $g$, in Eq.~(\ref{original_sto_sch_equ}). The number of flavor for light quarks is $N_{f}=3$, and $T$ denotes the temperature of the medium. In Eq.~(\ref{A field_expand}), $\vec{p}$ is the momentum of the heavy quark, $d_{g}=16$ is the color-spin degeneracy of the gluon. The authors~\cite{wupreparing} introduced a time-dependent random phase factor $e^{i \theta(\vec{p},t)}$ for each momentum mode of the gluon field. These random phase factors parameterize the thermal fluctuation of the gluon field on the amplitude level.

The matrix element of the interaction Hamiltonian in momentum basis takes the following form:
\begin{equation}
    \begin{split}
        \bra{\vec{p},s}H_{I}\ket{\vec{p}\,',s'}&=\int \frac{d^{3}\vec{x}gA(\vec{x})}{\sqrt{2E_{q,\vec{p}}}\sqrt{2E_{q,\vec{p}\,'}}} \bar{u}^{s}(p)\gamma^{0}u^{s'}(p')\\
        & \quad e^{-i(\vec{p}\,'-\vec{p})\cdot \vec{x}}e^{i\theta(\vec{p}\,'-\vec{p},t)}\\
        &\approx g\sqrt{\frac{2d_{g}}{E_{g}}} a(\vec{p}\,'-\vec{p}) \delta^{s's},\\
          E_{g}&=\sqrt{(\vec{p}\,'-\vec{p})^{2}+m_{g}^{2}}.
    \end{split}
\label{eq:hamiltonian_derivation}
\end{equation}
Here $E_{q,\vec{p}}=\sqrt{\vec{p}^{2}+m_{q}^{2}}$ is the kinetic energy of the heavy quark. $p=(E_{q,\vec{p}},\vec{p})$ is its 4-momentum. In this derivation, the following non-relativistic approximations are adopted
\begin{equation}
\begin{split}
    \bar{u}^{s}(p)\gamma^{0}u^{s'}(p')&\approx 2E_{q,\vec{p}}\delta^{s's}, \\
    \vec{p}&\approx \vec{p}\,',\\
    E_{q,\vec{p}}&\approx m_{q}.
\end{split}
\label{eq:non-relativistic_approx}
\end{equation}
Since the authors in~\cite{wupreparing} ignored the spin degrees of freedom, the $\delta^{s's}$ factor is omitted in the stochastic Schr\"{o}dinger equation.
   
After substituting Eq.~(\ref{eq:hamiltonian_derivation}) into Eq.~(\ref{original_sto_sch_equ}), we obtain the relativistic Schr\"{o}dinger equation in momentum space describing the interaction between the heavy quark and the thermal gluon field,
  \begin{equation}
  \label{sto_sch_equ}
  \begin{split}
  i \frac{\partial}{\partial t} \varphi(\vec{p},t)&=\sqrt{\vec{p}^{2}+m_{q}^{2}} \varphi(\vec{p},t) \\
  &+ g\int d^{3}\vec{p}\,' \sqrt{\frac{2d_{g}}{E_{g}}} a(\vec{p}\,'-\vec{p}) \varphi(\vec{p}\,',t).
  \end{split}
  \end{equation}
Here $\varphi(\vec{p},t)$ is the wave function of the heavy quark in momentum space, and $\sqrt{\vec{p}^{2}+m_{q}^{2}}$ is the heavy quark's kinetic energy.

Since the main goal of this paper is to illustrate the connection between the SSE and Boltzmann equation, for simplicity, we neglect the dissipation term in the interaction term of Eq.~(\ref{sto_sch_equ}), and  as a result, this equation will lead to a uniform distribution in momentum space as the equilibrium limit for the heavy quark~\cite{wupreparing}. For the SSE with the dissipation term, we refer to~\cite{wupreparing}, where the Boltzmann distribution is reached as the equilibrium distribution for the heavy quark.

Alternatively, Eq.~(\ref{sto_sch_equ}) can be rewritten in the form of the Lippmann-Schwinger equation~\cite{weinberg1995quantum}:
\begin{equation}
\begin{split}
    \ket{\varphi}&=\ket{\varphi_{0}}+H_{I}\tilde{G}^{0}_{F} \ket{\varphi}.
\end{split}
\end{equation}
Here $\tilde{G}^{0}_{F}=\frac{1}{E-H_{0}+ i\epsilon}$ is the noninteracting Green's function for the heavy quark at zero temperature. $\ket{\varphi_{0}}$ denotes the wave function for the free heavy quark. $H_{0}$ ($H_I$) is the kinetic energy (interaction) operator of the SSE, see Eq~(\ref{original_sto_sch_equ}). $E$ is the energy of the heavy quark. The corresponding interacting Green's function satisfies the following relation, 
\begin{equation}
\tilde{G}_{F}=\tilde{G}^{0}_{F}+\tilde{G}^{0}_{F}H_{I}\tilde{G}^{0}_{F}+\tilde{G}^{0}_{F}H_{I}\tilde{G}^{0}_{F}H_{I}\tilde{G}^{0}_{F}+....,
\label{Lippmann_Schwinger}
\end{equation}
where $\tilde{G}_{F}$ denotes the interacting Green's function for the heavy quark at zero-temperature.

\begin{figure*}[htbp]
  \centering
  \begin{tikzpicture}
  \begin{feynman}
  \vertex (a1);
  \vertex[right=2.0cm of a1](a2);
  \vertex[right=5.0cm of a1](a3);
  \vertex[right=9.0cm of a1](a4);
  \vertex[right=1.0cm of a1] (b1);
    \vertex[right=1.0cm of a2] (b2);
      \vertex[right=1.0cm of a3] (b3);
        \vertex[right=1.0cm of a4] (b4);
  \vertex[right=1.0cm of b2] (c2);
    \vertex[right=1.0cm of b3] (c3);
      \vertex[right=1.0cm of b4] (c4);
  \vertex[right=1.0cm of c3] (d3);
    \vertex[right=1.0cm of c4] (d4);
  \vertex[right=1.0cm of d4] (e4);
  \vertex[above=1.0cm of b2] (z2);
    \vertex[above=1.0cm of b3] (z3);
      \vertex[above=1.0cm of b4] (z4);
  \vertex[above=1.0cm of c3] (y3);
    \vertex[above=1.0cm of c4] (y4);
  \vertex[above=1.0cm of d4] (x4);
  \diagram* {
  (a1) -- [fermion] (b1),
  };
  \node at (0.0,-0.5) {$\sigma$};
  \node at (1.0,-0.5) {$\rho$};
  \node at (1.5,0) {+};
  \diagram* {
  (a2) -- [fermion] (b2) -- [fermion] (c2),
  {(z2) -- [insertion=0, charged scalar, momentum=\(p_{1}\)] (b2)},
  };
  \node at (2.0,-0.5) {$\sigma$};
  \node at (3.0,-0.5) {$\alpha$};
  \node at (4.0,-0.5) {$\rho$};
  \node at (4.5,0) {+};
  \diagram* {
  (a3) -- [fermion] (b3) -- [fermion] (c3) -- [fermion] (d3),
  {(z3) -- [insertion=0, charged scalar, momentum=\(p_{1}\)] (b3)},
  {(y3) -- [insertion=0, charged scalar, momentum=\(p_{2}\)] (c3)},
  };
  \node at (5.0,-0.5) {$\sigma$};
  \node at (6.0,-0.5) {$\alpha$};
  \node at (7.0,-0.5) {$\beta$};
  \node at (8.0,-0.5) {$\rho$};
  \node at (8.5,0) {+};
  \diagram* {
  (a4) -- [fermion] (b4) -- [fermion] (c4) -- [fermion] (d4) -- [fermion] (e4),
  {(z4) -- [insertion=0,charged scalar, momentum=\(p_{1}\)] (b4)},
  {(y4) -- [insertion=0, charged scalar, momentum=\(p_{2}\)] (c4)},
  {(x4) -- [insertion=0, charged scalar, momentum=\(p_{3}\)] (d4)},
  };
  \node at (9.0,-0.5) {$\sigma$};
  \node at (10.0,-0.5) {$\alpha$};
  \node at (11.0,-0.5) {$\gamma$};
  \node at (12.0,-0.5) {$\beta$};
  \node at (13.0,-0.5) {$\rho$};
  \node at (13.5,0) {+ ...};
  \end{feynman}
  \end{tikzpicture}
  \caption{The perturbative expansion of the heavy quark Green's function $G^{\rho\sigma}_{F}$ with an external classical gluon field $A$ coupled to the heavy quark. The dashed lines represent the external $A$ field in Eq.~(\ref{A field_expand}). The solid lines represent the heavy quark Green's functions. $\alpha, \beta, \gamma, \rho, \sigma = +, -$ are the labels of different kinds of the Keldysh Green's functions in Eq.~(\ref{keldysh_green_function}).}
    \label{external_field_pic}
  \end{figure*}
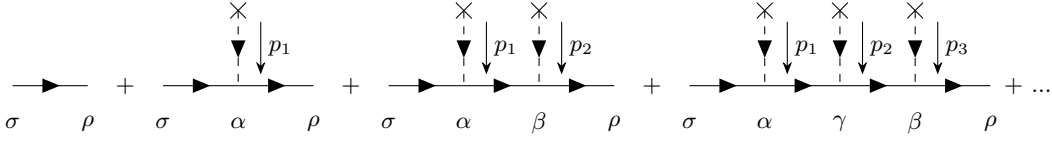

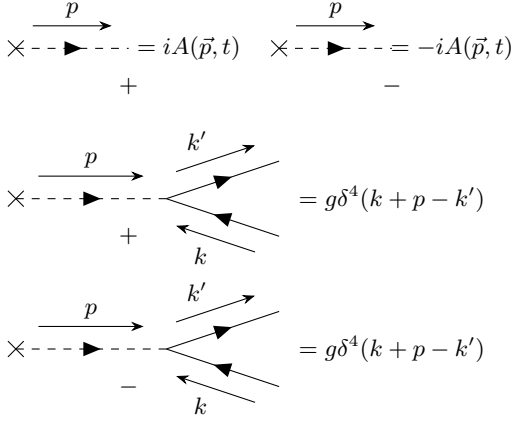
\begin{figure}
     \centering
  \begin{tikzpicture}
  \begin{feynman}
\vertex (a1);
\vertex[right=1.5cm of a1] (b1);
\vertex[right=3.5cm of a1] (a2);
\vertex[right=1.5cm of a2] (b2);
\vertex[below=2.0cm of a1] (a3);
\vertex[right=2.0cm of a3] (b3);
\vertex[below=1.5cm of a2] (c1);
\vertex[below=1.0cm of c1] (c2);
\vertex[below=2.0cm of a3] (a4);
\vertex[below=2.0cm of b3] (b4);
\vertex[below=2.0cm of c1] (c3);
\vertex[below=2.0cm of c2] (c4);
 \diagram* {
  (a1) -- [insertion=0, charged scalar, momentum=\(p\)] (b1),
  };
  \node at (1.5,-0.5) {$+$};
  \node at (2.3,0.0) {$=iA(\vec{p},t)$};
\diagram* {
  (a2) -- [insertion=0, charged scalar, momentum=\(p\)] (b2),
  };
  \node at (5.0,-0.5) {$-$};
  \node at (5.8,0.0) {$=-iA(\vec{p},t)$};
  \diagram* {
  (a3) -- [insertion=0, charged scalar, momentum=\(p\)] (b3),
  {(b3) -- [fermion, momentum=\(k'\)] (c1)},
  {(c2) -- [fermion, momentum=\(k\)] (b3)},
  };
  \node at (1.5,-2.5) {$+$};
  \node at (5.0,-2.0) {$=g\delta^{4}(k+p-k')$};
\diagram* {
  (a4) -- [insertion=0, charged scalar, momentum=\(p\)] (b4),
  {(b4) -- [fermion, momentum=\(k'\)] (c3)},
  {(c4) -- [fermion, momentum=\(k\)] (b4)},
  };
  \node at (1.5,-4.5) {$-$};
  \node at (5.0,-4.0) {$=g\delta^{4}(k+p-k')$};
  \end{feynman}
  \end{tikzpicture}
    \caption{Feynman rules for the external classical $A$ field, which is denoted by the cross~\cite{landau1981course}.}
    \label{fig:element_feynman_physical_external}
\end{figure}

Here, we assume the ensemble average of the zero-temperature Green's function corresponding to the SSE is equal to the Keldysh Green's function, that is,
\begin{equation}
\langle \tilde{G}_{\mathrm{SSE}}\rangle =G_{\mathrm{Keldysh}}.
\label{G_SSE_G_BE}
\end{equation}
Here $\langle  \hat{O}\rangle $ denotes the ensemble average of the operator $\hat{O}$~\cite{wupreparing}, which is taken for all observables evaluated from the SSE. When the coupling constant is small enough, the contributions from higher-order terms in Eq.~(\ref{Lippmann_Schwinger}) are expected to be small, so $\langle \tilde{G}_{\mathrm{SSE}}\rangle $ can be expanded perturbatively, as illustrated in Fig.~(\ref{external_field_pic})~\cite{landau1981course}.  The Feynman rules for the external field are illustrated in Fig.~(\ref{fig:element_feynman_physical_external}). The dashed lines represent the $A$ field and are labeled with $\pm$ at the interaction vertices.

In order to derive the Boltzmann equation from the stochastic Schr\"{o}dinger equation, let us first work out the corresponding gluon distribution from Eq.~(\ref{sto_sch_equ}).

The random phase factor in the gluon field satisfies $e^{i \theta(-\vec{p},t)}=e^{-i\theta(\vec{p},t)}$~\cite{wupreparing}, which leads to a real $A(\vec{x},t)$ field,
\begin{equation}
\label{A field_momentum}
\begin{split}
A(\vec{p},t)&=\sqrt{\frac{2d_{g}}{E_{\vec{p}}}}a(\vec{p}, t)=\sqrt{\frac{2d_{g}}{E_{\vec{p}}}} \sqrt{n(\vec{p})} e^{i\theta(\vec{p},t)}, \\
A(-\vec{p},t)&=\sqrt{\frac{2d_{g}}{E_{-\vec{p}}}}a(-\vec{p},t)=\sqrt{\frac{2d_{g}}{E_{-\vec{p}}}}\sqrt{n(-\vec{p})}e^{i\theta(-\vec{p},t)} \\
&=\sqrt{\frac{2d_{g}}{E_{\vec{p}}}}\sqrt{n(\vec{p})}e^{-i\theta(\vec{p},t)}\\
&=A^{*}(\vec{p},t).
\end{split}
\end{equation}   
Here $A^{*}(\vec{p},t)$ is the complex conjugate of $A(\vec{p},t)$. Therefore, $A(\vec{x},t)$ is real.

In~\cite{wupreparing}, the random phases are assumed to be uncorrelated for different momentum $\vec{p}$, and correlated within the time interval of $t_{\mathrm{corr}}$, that is, the random phases are updated every period of the correlation time $t_{\mathrm{corr}}$. $t_{\mathrm{corr}}$ is assumed to be independent of the momentum $\vec{p}$. The exact time when the phases get updated for different momentum is also randomized such that the random phases for different momenta $\vec{p}$ are updated asynchronously, see~\cite{wupreparing} for the details of the numerical implementation of $t_{\mathrm{corr}}$. Later in this section, we will show that the time correlation of the random phases is connected to the energy distribution of the gluons in the medium.
 
Following Eq.~(\ref{G_SSE_G_BE}), we now consider the ensemble average on the $A$ field and its correlations, which will be needed for constructing the scattering term of the Boltzmann equation from the SSE. In Fig.~(\ref{external_field_pic}), the one-vertex term vanishes due to $\langle A(\vec{p},t)\rangle \propto\langle e^{i\theta(\vec{p},t)}\rangle =0$ since the phase $\theta(\vec{p},t)$ is uniformly distributed over $\left[-\pi, \pi\right]$. Because when $\vec{p}_{1}\neq\vec{p}_{2}$, $\theta(\vec{p}_{1},t)$ is independent of $\theta(\vec{p}_{2},t')$,  the functions of two phases of different momenta are also independent of each other. Therefore, the $A$ field satisfies the following relation,
  \begin{equation}
  \begin{split}
  &\langle A^{*}(\vec{p}_{1},t_{1})A(\vec{p}_{2},t_{2})\rangle  \\
  &=\sqrt{\frac{2d_{g}}{E_{\vec{p}_{1}}}}\sqrt{\frac{2d_{g}}{E_{\vec{p}_{2}}}}\langle a^{*}(\vec{p}_{1},t_{1})a(\vec{p}_{2},t_{2})\rangle \\
  &=\sqrt{\frac{2d_{g}}{E_{\vec{p}_{1}}}}\sqrt{\frac{2d_{g}}{E_{\vec{p}_{2}}}}\sqrt{n(\vec{p}_{1})}\sqrt{n(\vec{p}_{2})}\langle e^{-i\theta(\vec{p}_{1},t_{1})}e^{i\theta(\vec{p}_{2},t_{2})}\rangle \\
  &=\sqrt{\frac{2d_{g}}{E_{\vec{p}_{1}}}}\sqrt{\frac{2d_{g}}{E_{\vec{p}_{2}}}}\sqrt{n(\vec{p}_{1})n(\vec{p}_{2})}\delta(\vec{p}_{1}-\vec{p}_{2}) \\
  &\quad \times \langle e^{-i\theta(\vec{p}_{1},t_{1})}e^{i\theta(\vec{p}_{2},t_{2})}\rangle .
  \end{split}
  \label{A field_corr_physical}
  \end{equation} 
When $|t_{1}-t_{2}|\leq t_{\mathrm{corr}}$, the correlation of the $A$ fields at two different times but the same momentum is nonzero. Therefore, the two-vertex terms are the lowest-order nonvanishing terms in the expansion of $\langle G_{\mathrm{SSE}}\rangle $ in Eq.~(\ref{Lippmann_Schwinger}). Here, we note that the correlation of the $A$ field with different momenta vanishes~\cite{wupreparing}.

Because the correlation function of the random phases is determined only by the difference in time and momentum, the correlation of $e^{i\theta(\vec{p},t)}$ is also determined only by these differences. In order to study the time correlation of the $A$ field, we define the following correlation function of the random phase factor $e^{i\theta(\vec{p},t)}$ as,
\begin{equation}
    g(\vec{p},\Delta t)=\frac{1}{T_{\mathrm{max}}}\int^{T_{\mathrm{max}}}_{0}\langle e^{-i\theta(\vec{p},t+\Delta t)}e^{i\theta(\vec{p},t)}\rangle  dt,
\label{define_of_gdeltat}
\end{equation}
where $T_{\mathrm{max}}$ is the duration of the time evolution. In\cite{wupreparing} the SSE was solved numerically through stepwise evolution using the MSD algorithm~\cite{PhysRevE.49.4684}. If the phase $\theta(\vec{p}, t_{i})$ is known at each time step $t_i$, we could use the following discretized formula to obtain $g(\vec{p},\Delta t)$ numerically~\cite{gubner2006probability}
    \begin{equation}
  \label{E_distribution}
  g(\vec{p}, \Delta t)=\frac{1}{n-h}\langle \sum\limits_{i=1}^{n-h} e^{-i\theta(\vec{p},t_{i+h})}e^{i\theta(\vec{p},t_{i})}\rangle .
  \end{equation}
Here $n$ represents the total number of steps in an event, and $h=\frac{\Delta t}{t_{\mathrm{step}}}=\frac{(t_{i+h}-t_{i})}{t_{\mathrm{step}}}$, where $t_{\mathrm{step}}=t_{i+1}-t_{i}$ denotes the step length in time.
  
  Taking the complex conjugation of the correlation function Eq.~(\ref{define_of_gdeltat}), we can obtain $g^{*}(\vec{p},\Delta t)=g(\vec{p},-\Delta t)$ as,
  \begin{equation}
  \label{E_distribution_conj}
  \begin{split}
  g^{*}(\vec{p},\Delta t)&= \langle e^{-i\theta(\vec{p},t+\Delta t)}e^{i\theta(\vec{p},t)}\rangle ^{*} \\
  &=\langle e^{-i\theta(\vec{p},t)}e^{i\theta(\vec{p}, t+\Delta t)}\rangle =g(\vec{p},-\Delta t).
  \end{split}
  \end{equation}
This property will later be used to obtain the energy distribution of the gluon field.

Since the random phases $\theta(\vec{p},t)$ are independently sampled for each different momentum $\vec{p}$, its time correlation $g(\vec{p},\Delta t)$ is independent of $\vec{p}$. So from now on, we will write $g(\vec{p},\Delta t)$ as $g(\Delta t)$ for simplicity.

In terms of $g(\Delta t)$, the correlation of the $A$ field can be written as
    \begin{equation}
  \langle A^{*}(\vec{p}_{1},t_{1})A(\vec{p}_{2},t_{2})\rangle =\frac{2d_{g}}{E_{\vec{p}_{1}}}n(\vec{p}_{1}) \delta(\vec{p}_{1}-\vec{p}_{2})g(\Delta t).
  \label{A field_corr_gt}
  \end{equation}
 
To obtain the corresponding Boltzmann equation from the SSE, we need to know the Wigner distribution of the gluons in the thermal medium, which can be derived from the Wigner transformation of the correlation of the $A$ field. The resulting $W_{A}$ is
\begin{equation}
\begin{split}
&W_{A}(\vec{X},\vec{P},t,E) \\
&=\int \frac{d^{3}\vec{p}\,'}{(2\pi)^{4}} dt' \langle A^{*}(\vec{P}-\frac{\vec{p}\,'}{2}, t+\frac{t'}{2})A(\vec{P}+\frac{\vec{p}\,'}{2}, t-\frac{t'}{2})\rangle  \\
& \quad \times e^{i\vec{X}\cdot\vec{p}\,'}e^{iEt'} \\
&=\int \frac{d^{3}\vec{p}\,'}{(2\pi)^{4}} dt' \frac{2d_{g}}{E_{\vec{P}-\frac{\vec{p}\,'}{2}}}n(\vec{P}-\frac{\vec{p}\,'}{2})\delta^{3}(\vec{p}\,')g(t')e^{i\vec{X}\cdot\vec{p}\,'}e^{iEt'}\\
&=\int \frac{dt'}{(2\pi)^{4}} \frac{2d_{g}}{E_{\vec{P}}} n(\vec{P})g(t')e^{iEt'}\\
&=\frac{2d_{g}}{(2\pi)^4 E_{\vec{P}}} n(\vec{P}) G(E).
\end{split}
\label{wigner_A}
\end{equation}    
Here we assume that the thermal field follows a uniform distribution in position $\vec{X}$ and time $t$, therefore $g(t')\approx \langle e^{-i\theta(\vec{p},t+\frac{t'}{2})}e^{i\theta(\vec{p},t-\frac{t'}{2})}\rangle $. In the last equality, $G(E)$ is the Fourier transformation of $g(\Delta t)$,
\begin{equation}
    G(E)=\int d\Delta t g(\Delta t)e^{i\Delta tE}.
\label{G(E)_define}
\end{equation}
By comparing the expression of the Keldysh Green's function for the boson in Eq.~(\ref{G-+}), and Eq.~(\ref{wigner_A}), we find that the correlation function of the $A$ field resembles the Keldysh Green's function $G^{-+}_{B}$, and that $W_{A}(\vec{X},\vec{P},t,E)$ resembles the Wigner transformation of $G_{B}^{-+}$. The only difference is on the energy dependence: in Eq.~(\ref{G-+}), the boson's energy is on-shell; whereas in $W_A(\vec{X},\vec{P},t,E)$, its energy dependence is given by $G(E)$. If we assume that the Keldysh Green's functions for the $A$ field can be constructed in terms of its correlation functions $G^{-+}_{A}(x,y)=-i\langle A^{*}(y)A(x)\rangle $ and $ G^{+-}_{A}(x,y)=-i\langle A(x)A^{*}(y)\rangle $, we can obtain the Wigner transformation on $iG^{-+}_{A}$ and $iG_A^{+-}$ as,
\begin{equation}
\begin{split}
\mathcal{W} \left[iG^{-+}_{A} \right]&=iG^{-+}_{A,w}(X,P)=W_{A}(\vec{X},\vec{P},t,E), \\
\mathcal{W} \left[iG^{+-}_{A} \right]&=iG^{+-}_{A,w}(X,P)=W_{A}(\vec{X},\vec{P},t,E).
\end{split}
\label{wigner_A_pm_mp}
\end{equation}
Here $X=(t,\vec{X}), P=(E,\vec{P})$. We note that in the approximation of treating the gluon field as a classical field, the Wigner transformation of $iG^{-+}_A$ and $iG^{+-}_A$ is identical. By comparing the $A$ field's correlation function $W_A(\vec{X},\vec{P},t,E)$ and the Fourier expansion of the boson's Keldysh Green's function $G_{B}$, one can identify $G(E)$ as the off-shell gluons' energy distribution. In order to ensure that the energy distribution function is real and nonnegative, $g(\Delta t)$ needs to be real and even. Now, we show that $g(\Delta t)$ indeed satisfies these properties.

We first show that $g(\Delta t)$ is a real function through the time correlation function of the random phases. Let us first consider the case when $|t'-t|\le t_{\mathrm{corr}}$. Since $\theta(\vec{p},t)$ and $\theta(\vec{p},t')$ are equal to each other when $|t'-t|\leq t_{\mathrm{corr}}$, and according to Eq.~(\ref{E_distribution}), we have
\begin{equation}
\begin{split}
    g(t_{i+h}-t_{i})&=g(\vec{p},\Delta t)\\
    &=\frac{1}{n-h}\langle \sum\limits_{i=1}^{n-h} e^{-i\theta(\vec{p},t_{i+h})}e^{i\theta(\vec{p},t_{i})}\rangle \\
    &=\frac{1}{n-h}\sum\limits_{i=1}^{n-h} \langle e^{-i\theta(\vec{p},t_{i})}e^{i\theta(\vec{p},t_{i})}\rangle .
\end{split}
\end{equation}
Here $\Delta t=t_{i+h}-t_{i}$, and note that in the last equality, the two phases are taken to be at the same time. Because $\langle e^{-i\theta(\vec{p},t)}e^{i\theta(\vec{p},t)}\rangle $ is a real function, $g(\Delta t)$ is also a real function. When $|t'-t|>  t_{\mathrm{corr}}$, these two phases are independent of each other. The ensemble average of their product is
\begin{equation}
\begin{split}
    &g(t-t')=g(\vec{p},\Delta t)=\\
    &\langle e^{-i\theta(\vec{p},t)}e^{i\theta(\vec{p},t')}\rangle =\langle e^{-i\theta(\vec{p},t)}\rangle \langle e^{i\theta(\vec{p},t')}\rangle =0,
\end{split}
\end{equation}
which is also real. Therefore, regardless of the value of $\Delta t$, $g(\Delta t)$ is a real function. Thus, according to Eq.~(\ref{E_distribution_conj}), $g(\Delta t)$ is an even function and therefore $G(E)$ is also an even function.

To establish the connection between the SSE and BE, we need to calculate the scattering term of BE from the SSE. Since we aim to derive BE from the SSE in the weak coupling limit, the dominant contribution comes from the term containing two $H_I$'s on the right-hand side of Eq.~(\ref{Lippmann_Schwinger}), which is represented by the two-vertex term in Fig.~(\ref{fig_green_function}). In BE, this term corresponds to the one-loop term in the series expansion of the heavy quark self-energy $\Sigma^{+-}_w$, which is represented by the first term in Fig.~(\ref{irreducible_parts}). Keeping only the one-loop term in $\Sigma^{+-}_w$, the second term of the right-hand side of Eq. (26) becomes:
\begin{equation}
\begin{split}
    &\Sigma^{+-}_{w}(X,P)G^{-+}_{F,w}(X,P) \\
    &=-i\Sigma^{+-}_{w}(X,P)\left[iG^{-+}_{F,w}(X,P)\right]\\
    &=g^{2}\int d^{3}\vec{k} dk^{0} \left[iG^{+-}_{A,w}(X,-k)\right]\left[iG^{+-}_{F,w}(X,P-k)\right] \\
    &\qquad\qquad\qquad\times\left[iG^{-+}_{F,w}(X,P)\right]\\
    &=\int d^{3}\vec{k} dk^{0}\frac{2g^{2}d_{g}}{(2\pi)^{4} E_{\vec{k}}}n(-\vec{k})G(-k^{0})\\
    &\quad \times \frac{1}{(2\pi)^{3}\left(2E_{q,\vec{P}-\vec{k}}\right)}\left[\gamma^{\mu}(P-k)_{\mu}+m_{q}\right]\left[1-f(X,P-k)\right]\\
    &\quad \times \frac{1}{(2\pi)^{3}\left(2E_{q,\vec{P}}\right)}\left(\gamma^{\nu}P_{\nu}+m_{q}\right)\left[-f(X,P)\right]\\
    &\approx -g^{2}\int d^{3}\vec{k} dk^{0} \left[\gamma^{\mu}(P-k)_{\mu}+m_{q}\right]\left(\gamma^{\nu}P_{\nu}+m_{q}\right)\\
    &\quad \times \frac{d_{g}} {2(2\pi)^{10}E_{\vec{k}}E_{q,\vec{P}-\vec{k}}E_{q,\vec{P}}}n(\vec{k})G(k^{0})f(X,P).
\end{split}
\label{selfenergy_greenfunction_simplify}
\end{equation}
Here $g$ is the coupling constant, $P^{\mu}=(E_{q,\vec{P}},\vec{P})$ and $ k^{\mu}=(k^{0},\vec{k})$ are 4-momentum of the heavy quark and the gluon respectively. $E_{q,\vec{P}}$ is the kinetic energy of the heavy quark with the momentum $\vec{P}$, and satisfies the mass-shell relation which comes from the Wigner transformation of $G^{-+}$ in Eq.~(\ref{wigner_Green's_function}). In the third equality, we substitute $G^{+-}_{F,w}$ with the Wigner transformation of the Green's functions in Eq.~(\ref{Wigner_G+-_F0}) and the gluon Keldysh Green's function with the transformation of the correlation of $A$ field in Eq.~(\ref{wigner_A}). In the final step, we ignore the Pauli blocking factor $1-f(X,P-k)$ because the system we consider contains only one heavy quark. By analogy with Eq.~(\ref{selfenergy_greenfunction_simplify}), we obtain the result of $\Sigma^{-+}_{w}(X,P)G^{+-}_{F,w}(X,P)$ as,
\begin{equation}
\begin{split}
    &\Sigma^{-+}_{w}(X,P)G^{+-}_{F,w}(X,P) \\
    &\approx -g^{2} \int d^{3}\vec{k} dk^{0}\left(\gamma^{\nu}P_{\nu}+m_{q}\right)\left[\gamma^{\mu}(P-k)_{\mu}+m_{q}\right] \\
    &\quad \times \frac{d_{g}}{2(2\pi)^{10}E_{\vec{k}}E_{q,\vec{P}-\vec{k}}E_{q,\vec{P}}}n(\vec{k})G(k^{0})f(X,P-k).
\end{split}
\label{selfenergy_greenfunction_simplify_2}
\end{equation}

Next we sum over the spin degrees of freedom of the heavy quark by tracing out the spinor space on both sides of Eq.~(\ref{G_G_simplify}). The trace of the gamma matrices on the right-hand side of Eq.~(\ref{selfenergy_greenfunction_simplify_2}) leads to the squared scattering amplitude
      \begin{equation}
      \begin{split}
  |M|^{2}&=\frac{1}{4} g^{2}\mathrm{tr}\left\{ \left(\gamma^{\mu}P_{\mu}+m_{q}\right)\left[\gamma^{\nu}(P-k)_{\nu}+m_{q}\right]\right\}.
  \end{split}
  \label{noaverage_scattering_matrix}
  \end{equation}
Here the factor $\frac{1}{4}$ arises from tracing the drift term.
  
Therefore, the scattering process in the SSE can be interpreted as an on-shell heavy quark absorbing (or emitting) an off-shell gluon and forming a new on-shell heavy quark. Since the gluons are off-shell, there is no contradiction between momentum conservation and the heavy quark's mass-shell relation. Therefore the two-vertex term in Fig.~(\ref{external_field_pic}) does not vanish.

Due to the non-relativistic approximation adopted in the SSE, only the zeroth component of $\bar{u}(P)\gamma^{\mu}u(P')$ enters the interaction Hamiltonian~\cite{peskin2018introduction}. Therefore, the squared scattering amplitude can be further simplified as,
\begin{equation}
\begin{split}
|M|^{2}& \approx \frac{1}{4} g^{2} \mathrm{tr} \left\{ \left[ \bar{u}(P)\gamma^{0} u(P)\right] \left[\bar{u}(P)\gamma^{0} u(P) \right] \right\} \\
&=\frac{1}{4} g^{2} \mathrm{tr} \left( 4 m_{q}^{2} \mathbf{I} \right) \\
&= 4g^{2}m^{2}_{q}.
\end{split}
\label{scattering_matrix}
\end{equation}
Here $\mathbf{I}$ is the $4 \times 4$ identity matrix in the spinor space.

  In BE the energy and momentum in the scattering process are conserved, and they satisfy
  \begin{equation}
  \begin{split}
  \vec{P}-\vec{k}&=\vec{P}\,', \\
  \sqrt{\vec{P}^{2}+m_{q}^{2}}-k^{0}&=\sqrt{\vec{P}\,'^{2}+m_{q}^{2}},
  \end{split}
  \label{momentum_energy_conservation}
  \end{equation}
where $\vec{P}$ and $\vec{P}\,'$ are incoming and outgoing momentum of the heavy quark, respectively, and their corresponding kinetic energy are $\sqrt{\vec{P}^{2}+m_{q}^{2}}$ and $\sqrt{\vec{P}\,'^{2}+m_{q}^{2}}$ respectively. $\vec{k}$ and $k^{0}$ are the momentum and energy of the off-shell gluon participating in the scattering process, respectively.
  
Because $E_{q,\vec{P}}$ satisfies the mass-shell relation and only depends on the 3-momentum $\vec{P}$, the phase space distribution of heavy quarks depends only on the 3-momentum $\vec{P}$, 3-position $\vec{X}$, and time $t$. In the following part of this paper, we use $\tilde{f}(\vec{X},\vec{P},t)\equiv f(X,P)|_{P^0=E_{q,\vec{P}}}$ to denote the heavy quark's phase space distribution. For conciseness, we will drop the tilde on $\tilde{f}(\vec{X},\vec{P},t)$ from now on. By substituting the product of the self-energies and Green's functions into Eq.~(\ref{G_G_simplify}), performing the trace over the spinor space, setting $\vec{P}\,'=\vec{P}-\vec{k},E=k^{0}$ and integrating over $P^0$, we obtain the Boltzmann equation in the following form
    \begin{equation}
  \label{boltzmann_equation}
  \begin{split}
    &\frac{\partial f(\vec{X},\vec{P},t)}{\partial t}+\frac{\vec{P}}{E_{q,\vec{P}}} \cdot \frac{\partial f(\vec{X},\vec{P},t)}{\partial \vec{X}} = \frac{d_{g}}{2(2\pi)^{4}E_{q,\vec{P}}} \\
    &\times \int \frac{d^{3}\vec{P}\,'}{E_{q,\vec{P}\,'}}\int \frac{d^{3}\vec{k}}{E_{\vec{k}}} \int dE\, W'(\vec{P},\vec{P}\,',\vec{k},E_{q,\vec{P}},E_{q,\vec{P}\,'},E) \\
    &\quad \times \big[f(\vec{X},\vec{P}\,',t)n(\vec{k})G(E)-f(\vec{X},\vec{P},t)n(\vec{k})G(E)\big].
    \end{split}
  \end{equation}
Here $\vec{X}$ is the 3-position, $t$ is the time, $\vec{P}$ is the 3-momentum, and $\vec{k}$ denotes the momentum of the gluon absorbed or emitted by the heavy quark. $E_{q,\vec{P}}$ is the kinetic energy of the quark, and in the non-relativistic approximation, $E_{q,\vec{P}}\approx m_{q}$. The function $W'$ is related to the squared scattering amplitude $|M|^2$~\cite{van2015introduction, olive2014review} as,
    \begin{equation}
    \begin{split}
  &W'(\vec{P},\vec{P}\,',\vec{k},E_{q,\vec{P}},E_{q,\vec{P}\,'},E) \\
  &=\frac{|M|^{2}}{(2\pi)^{3}}(2\pi)^{4} \delta^{3}(\vec{P}-\vec{P}\,'-\vec{k}) \delta(E_{q,\vec{P}}-E_{q,\vec{P}\,'}-E).
  \end{split}
  \label{W_function}
  \end{equation}

In the above derivation, the gluon energy distribution $G(E)$ serves as a bridge connecting the SSE with the Boltzmann equation. It provides the energy distribution of the gluons in the Boltzmann equation following from the time dependence of  the random phases of the gluon field in the stochastic Schr\"{o}dinger equation. 

Let us now consider the equilibrium limit of the BE and SSE. Applying the Boltzmann's $H$-theorem~\cite{landau1981course} to the BE in Eq. (\ref{boltzmann_equation}), we find that the equilibrium distribution of the heavy quark in this system satisfies
\begin{equation}
f_{\mathrm{eq}}(\vec{X},\vec{P}\,',t)=f_{\mathrm{eq}}(\vec{X},\vec{P},t). 
\label{equilibrium_relation}
\end{equation}
This equation implies that the equilibrium distribution is uniform in momentum space (assuming a weak dependence on the position $X$), which arises from the absence of the dissipation term in the SSE considered in this paper, see the discussion below Eq.~(\ref{sto_sch_equ}). As shown in~\cite{wupreparing}, the SSE with the dissipation term included indeed leads to the Boltzmann distribution in the equilibrium limit~\cite{wupreparing} for the heavy quark.

Based on the above derivation, we demonstrate that in the weak coupling constant condition, the SSE in Eq.~(\ref{sto_sch_equ}) is consistent with the Boltzmann equation in Eq.~(\ref{boltzmann_equation}). In the next section, we will verify this consistency through numerical calculations.

\section{Numerical calculation}

In this section, we compare the evolution of the heavy quark distributions obtained from the SSE and the Boltzmann equation through numerical calculations. For demonstration purposes, we perform numerical calculations in 1+1 dimensional space-time.

In this section, we adopt the following values for the parameters: Our calculations are performed in discretized momentum space with the step size $\Delta p=0.0196\,\mathrm{GeV}$, where the momentum ranges in $\vec{p} \in \left[-\pi\,\mathrm{GeV},\pi\,\mathrm{GeV}\right]$. The coupling constant for the thermal mass of the gluon is $g_{0}=2.22$, and $\alpha_{s0}=\frac{g_{0}^{2}}{4\pi}=0.4,$ the medium temperature is $ T=0.15\,\mathrm{GeV}, $ the heavy quark mass is $m_{q}=1.0\,\mathrm{GeV}$, the gluon mass is $m_{g}=g_{0} \sqrt{1+\frac{N_{f}}{6}}T=0.41\,\mathrm{GeV}$ with the number of quark flavor $N_{f}=3$. Unless otherwise specified, the coupling constant between the heavy quark and the gluon field is $\alpha_{s}=\frac{g^{2}}{4\pi}=0.4, $ the correlation time for the gluon field is $t_{\mathrm{corr}}=0.067\,\mathrm{GeV}^{-1}$, and we take the ensemble average over $N_{\mathrm{event}}=100$ events for the numerical results shown in this section.

\begin{figure}[!h]
    \subfigure{
  \includegraphics[scale=0.12]{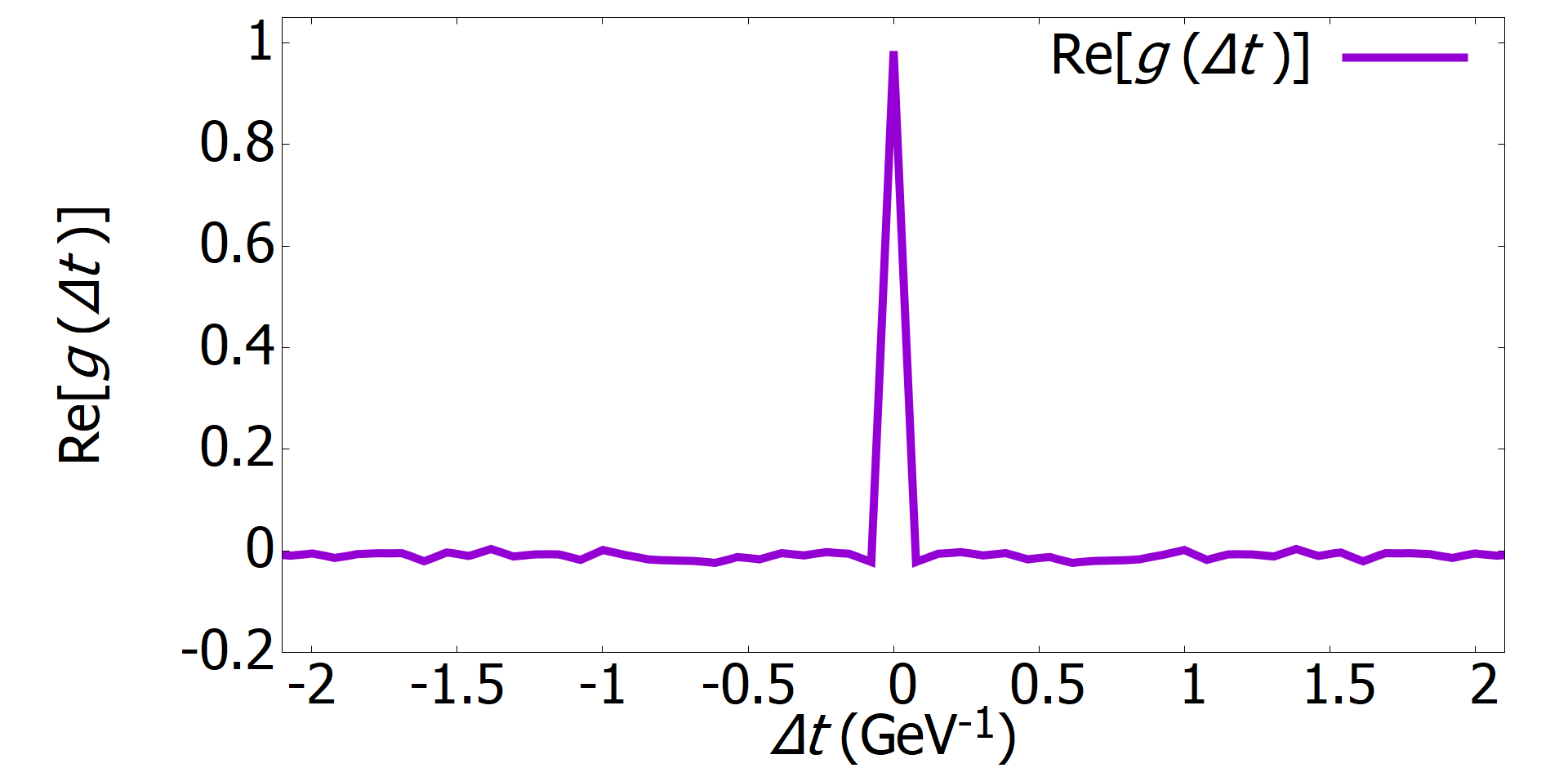} 
  }
     \subfigure{
  \includegraphics[scale=0.12]{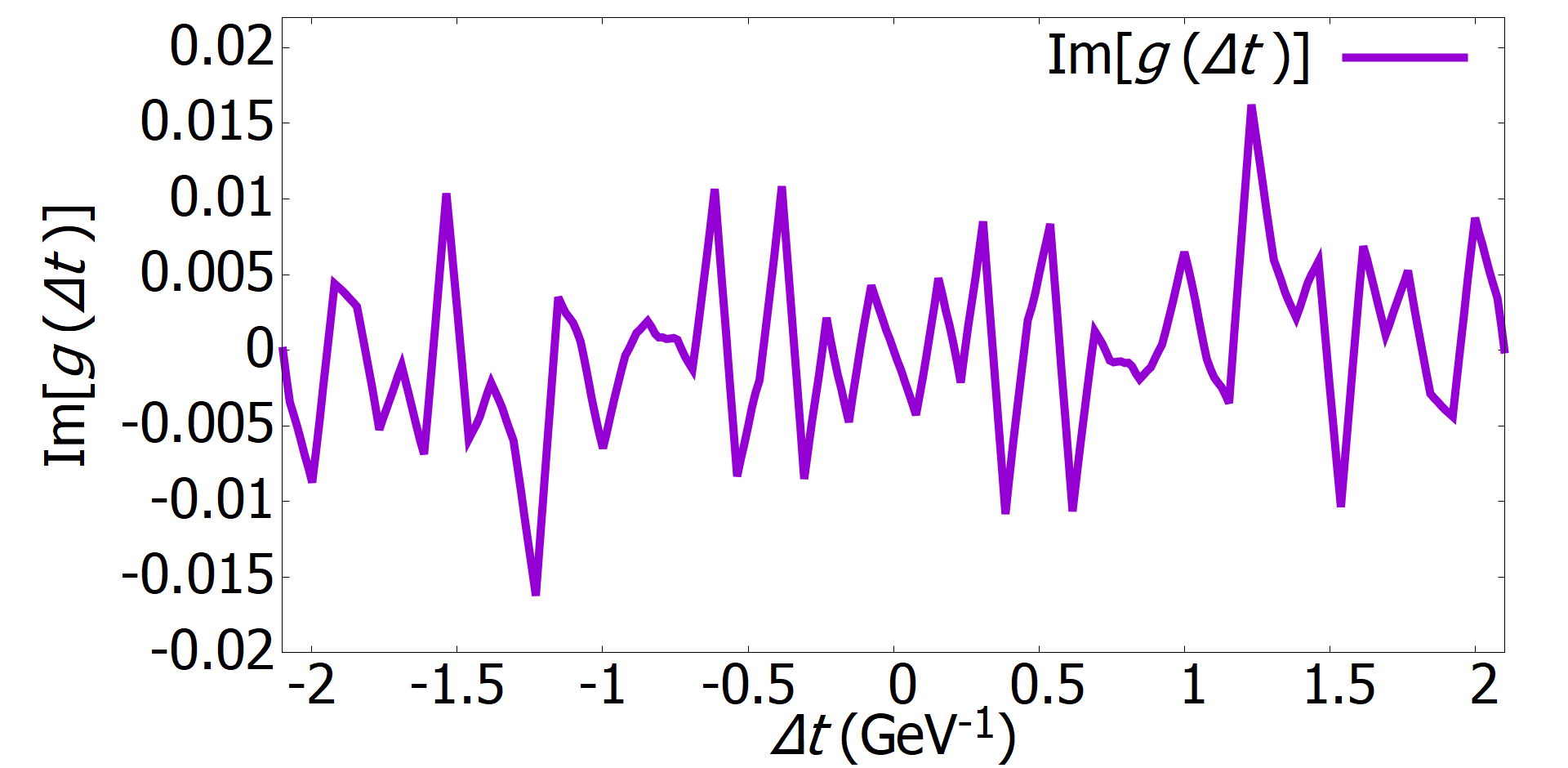}
  }
  \caption{The ensemble average of the time correlation function $g(\Delta t)$ of the random phases of the gluon field over 100 events, with $t_{\mathrm{corr}}=0.067\,\mathrm{GeV}^{-1}$. Top: real part of the correlation function $\Re\left[g(\Delta t) \right]$. Bottom: imaginary part of the correlation function $\Im\left[g(\Delta t) \right]$.}
  \label{fulltimeexpcorr_Re_Im}
  \end{figure}

Let us begin with analyzing the time correlation of the random phases. For each discretized momentum $\vec{p}_{i}$, the random sequences of the gluon phase $\theta_j(\vec{p}_i,t)$ of the 100 events are generated:
\begin{equation}
\{\theta_{j}(\vec{p}_{i},t)| t=0, t_{0}, 2t_{0}, ..., nt_{0}\}.
\label{random_series}
\end{equation}
Here $t$ is the time, $n$ is the number of time steps of each event, $j$ denotes the $j$th event, and $t_{0}$ is the time step size (much smaller than $t_{\mathrm{corr}}$). The corresponding sequences of $e^{i\theta(\vec{p}_{i},t)_{j}}$ read,
\begin{equation}
\{e^{i\theta_{j}(\vec{p}_{i},t)}| t=0, t_{0}, 2t_{0}, ..., nt_{0}\}.
\end{equation}
In each event, we resample the random phases after a period of $t_{\mathrm{corr}}$. In addition, we randomize the time when the phases are updated in each event, such that in different events, the random phases are updated at different times. Using Eq.~(\ref{E_distribution}), we can obtain the time correlation function $g_{j}(\Delta t)$ of $e^{i\theta_{j}(\vec{p}_{i},t)}$ for the $j$th event.
Then, the ensemble average of the time correlation functions is
\begin{equation}
  g(\Delta t)=\frac{1}{N_{\mathrm{event}}}\sum\limits_{j=1}^{N_{\mathrm{event}}} g_{j}(\Delta t) \quad (\Delta t \ge 0).
\end{equation}
Since $g^{*}(\Delta t)=g(-\Delta t)$, we can obtain the function $g(\Delta t<0)$ from $g(\Delta t> 0)$. The resulting $g(\Delta t)$ is shown in Fig.~(\ref{fulltimeexpcorr_Re_Im}).
As we can see from Fig.~(\ref{fulltimeexpcorr_Re_Im}), the real part of $g(\Delta t)$ reaches a maximum value of around 1.0 at $\Delta t=0$, and the correlation decreases as $|\Delta t|$ increases. When $|\Delta t| \ge t_{\mathrm{corr}}$, the correlation is close to 0. The imaginary part of $g(\Delta t)$ fluctuates around $\Im\left[g(\Delta t) \right]=0$, and its amplitudes are much smaller than $\Re\left[g(\Delta t) \right]$. Through numerical calculation, we find that $\mathrm{Im}[g(\Delta t)]$ tends to zero when the number of events increases, which confirms the fact that $G(E)$ is an even function.

\begin{figure}[!h]
    \subfigure{
  \includegraphics[scale=0.12]{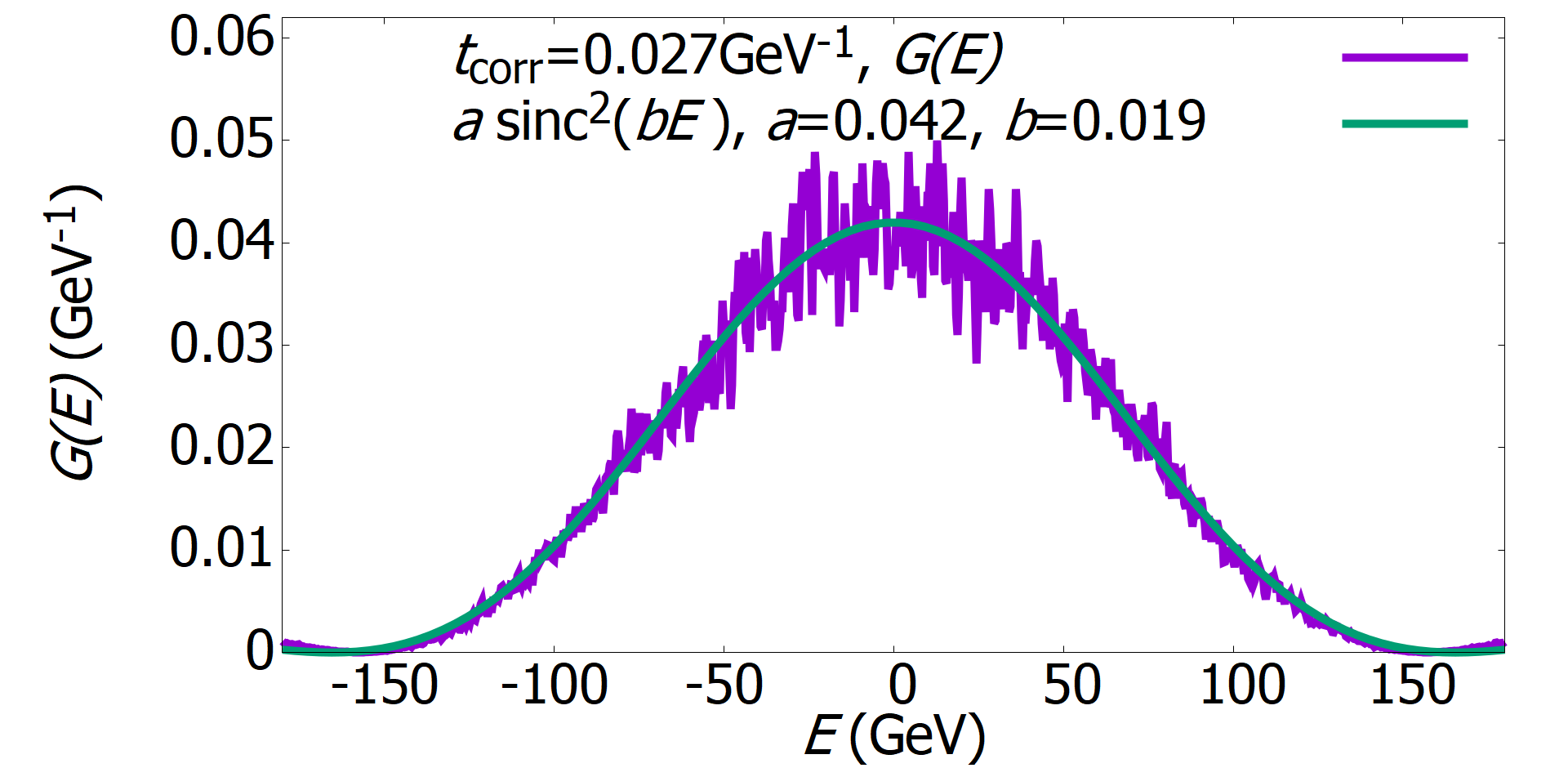} 
  }
  
  \subfigure{
  \includegraphics[scale=0.12]{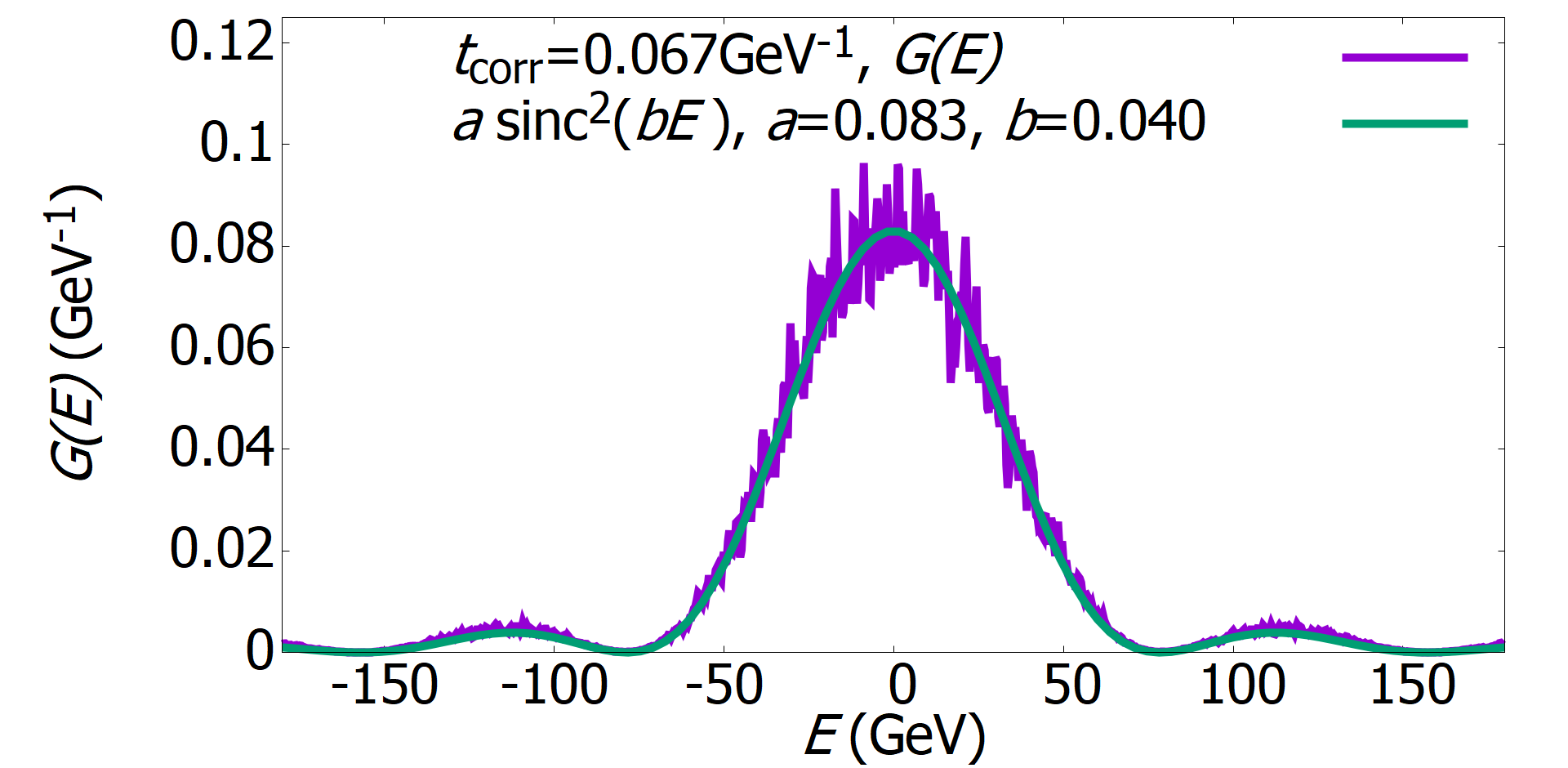} 
  }
  
  \subfigure{
  \includegraphics[scale=0.12]{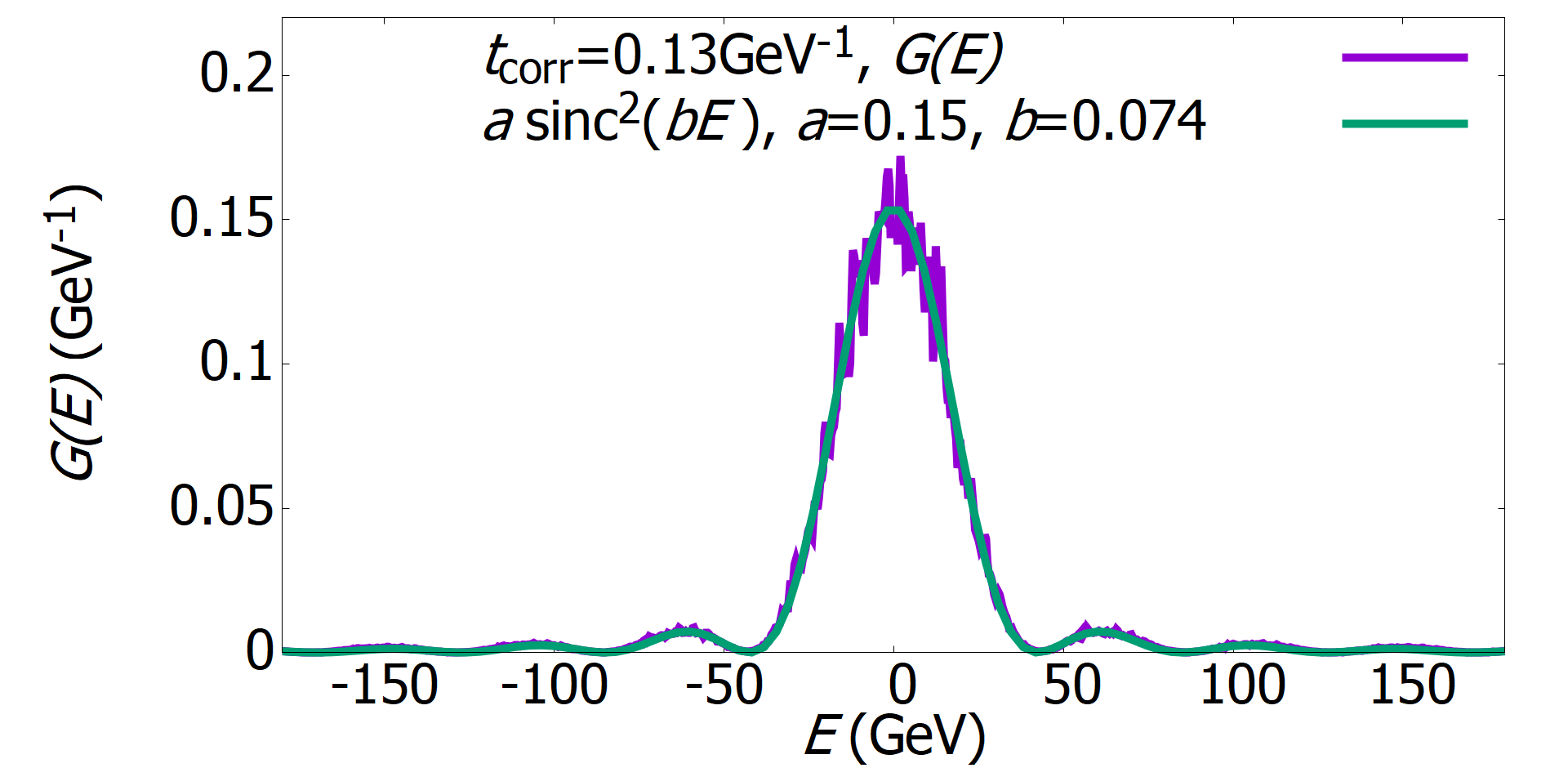} 
  }
  
  \subfigure{
  \includegraphics[scale=0.12]{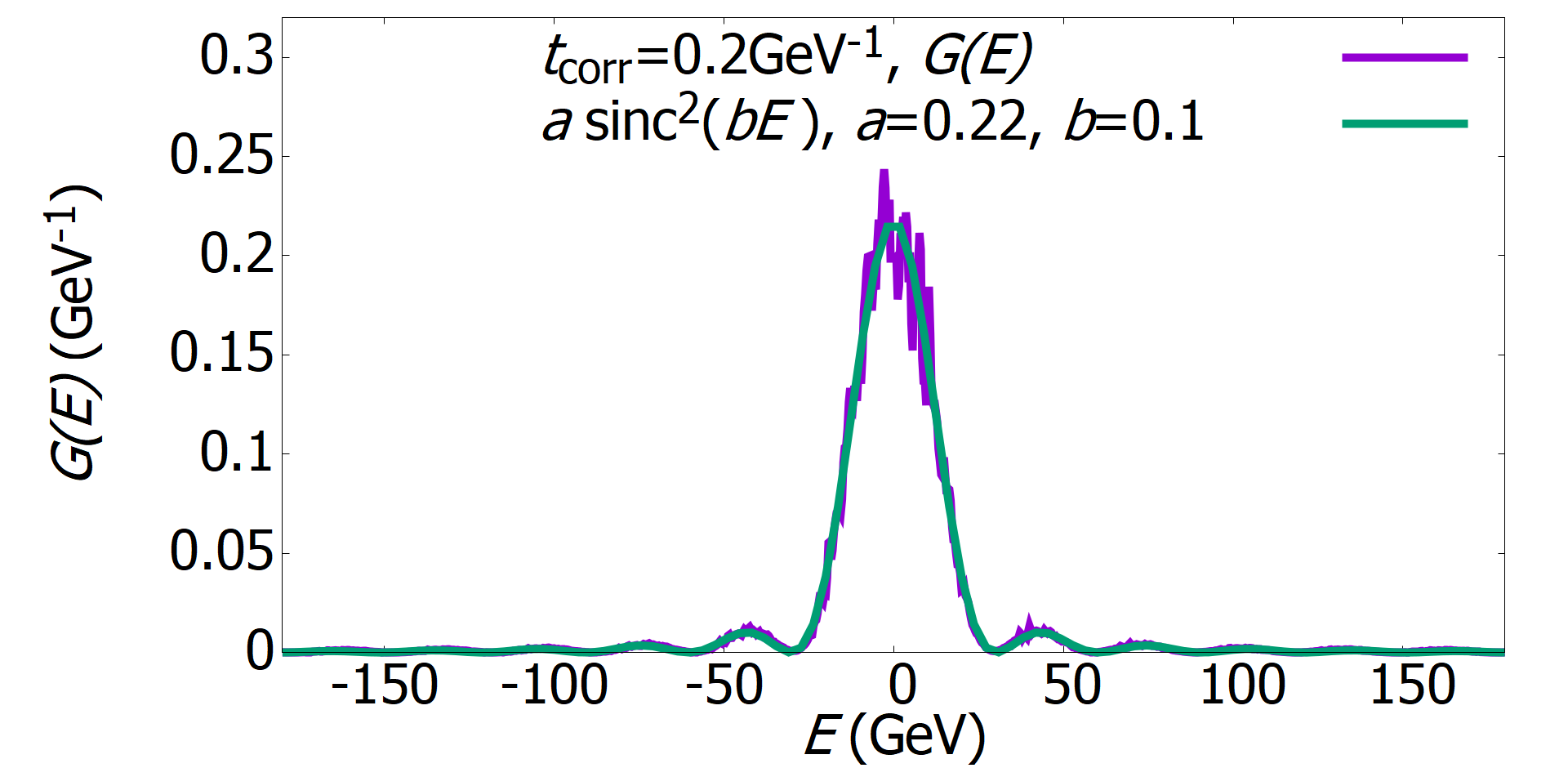} 
  }
  
  \caption{(Color online) The energy distribution of the gluon $G(E)$ from the SSE, with $t_{\mathrm{corr}}=0.027\,\mathrm{GeV}^{-1}, 0.067\,\mathrm{GeV}^{-1}, 0.13\,\mathrm{GeV}^{-1}, 0.2\,\mathrm{GeV}^{-1}$ from top panel to bottom panel. The unit of $a$ and $b$ in the legend is $\mathrm{GeV}^{-1}$. Purple line: Fourier transform of the time correlation function of the random phase factor $g(\Delta t)$; green line: fit result with the ansatz $G(E)=a\mathrm{sinc}^2(bE)$, see text for the detail.}
  \label{tcorrfunction}
  \end{figure}

\begin{figure}[!h]
      \subfigure{
  \includegraphics[scale=0.12]{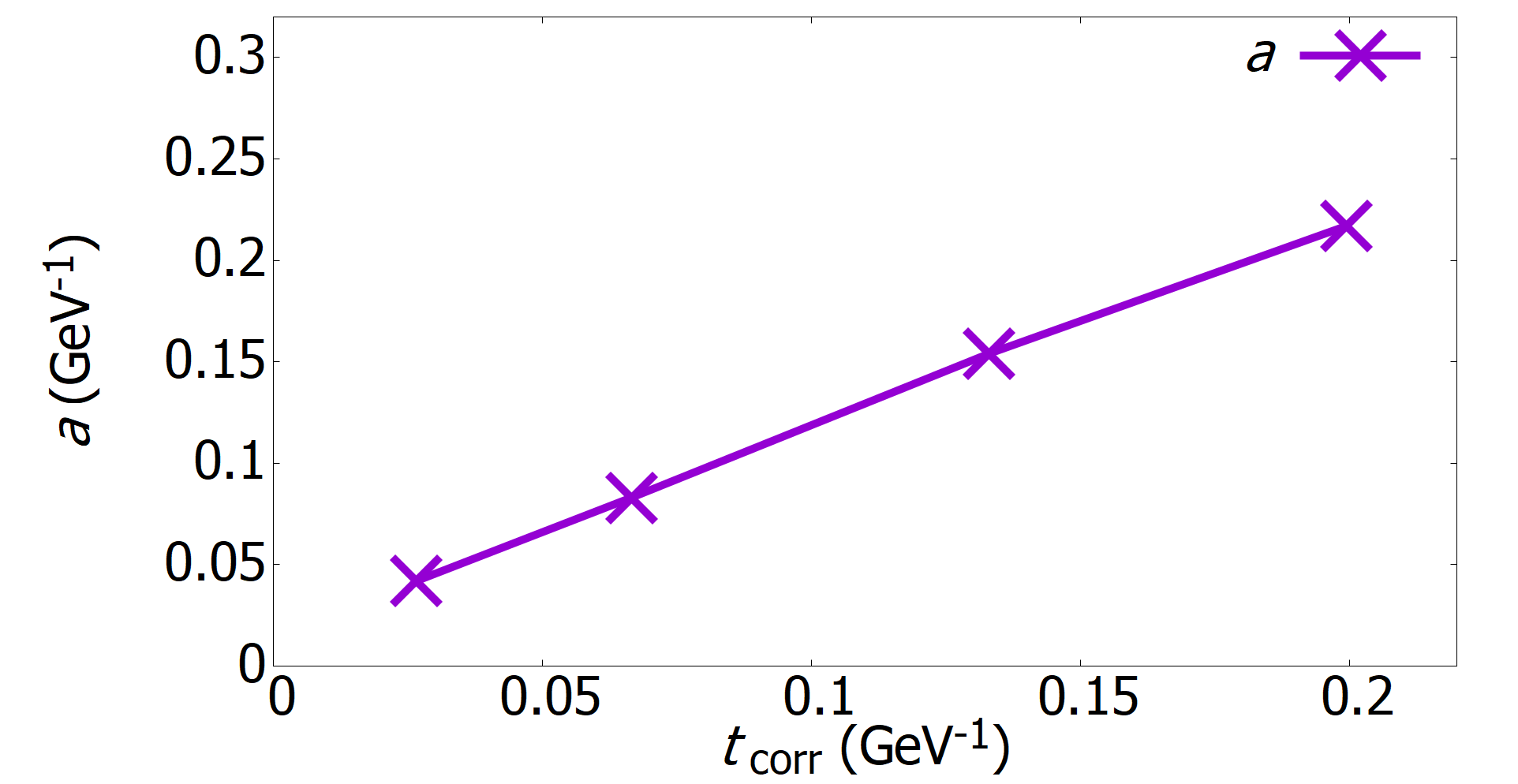} 
  }

  \subfigure{
  \includegraphics[scale=0.12]{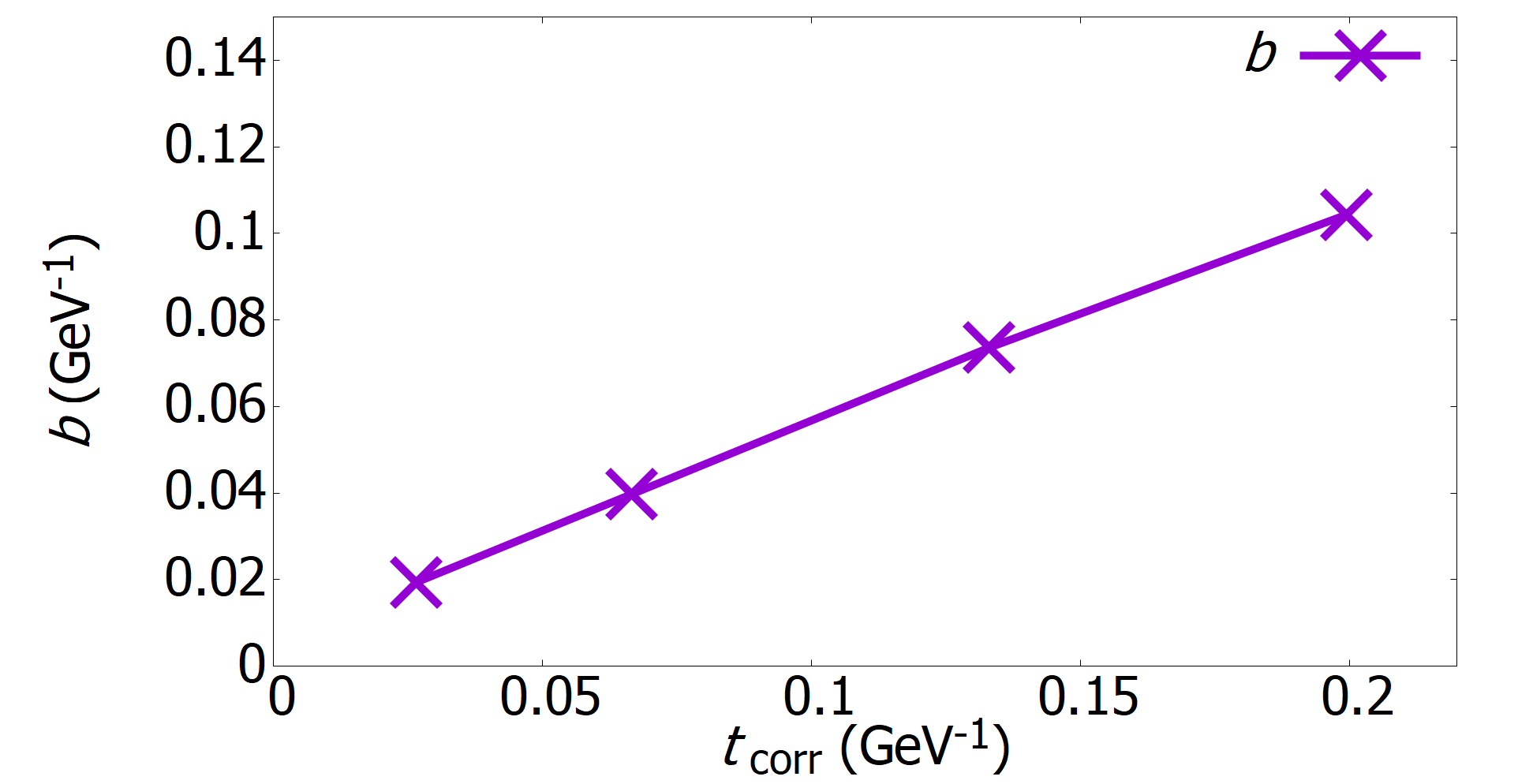} 
  }
  \caption{The relation between the fit parameters $a,b$ and the correlation time $t_{\mathrm{corr}}$, see text for the detail. Top panel: $a$ as a function of the correlation time $t_{\mathrm{corr}}$. Bottom panel: $b$ as a function of the correlation time $t_{\mathrm{corr}}$.}
  \label{a_b_tcorr}
  \end{figure}

The energy distribution of the gluon $G(E)$, as shown in Fig.~(\ref{tcorrfunction}), is the Fourier transform of the correlation function $g(\Delta t)$, see Eq.~(\ref{G(E)_define}). In Fig.~(\ref{tcorrfunction}), we use the ansatz $G(E)=a\mathrm{sinc}^{2}(bE)$ to fit the numerical results and determine the values of $a,b$ for different $t_{\mathrm{corr}}$. Through fitting the numerical results, we obtain 
\begin{equation}
\begin{split}
a&=1.0155t_{\mathrm{corr}}+0.0157\left(\mathrm{GeV}^{-1}\right),\\ b&=0.4899t_{\mathrm{corr}}+0.007\left(\mathrm{GeV}^{-1}\right).
\end{split}
\end{equation}
The parameters $a$ and $b$ are found to be proportional to $t_{\mathrm{corr}}$, as illustrated in Fig.~(\ref{a_b_tcorr}). As expected, the width of the energy distribution of the gluon $G(E)$ is inversely proportional to that in $g(\Delta t)$. By substituting the fitted $G(E)$ into our Boltzmann equation Eq.~(\ref{boltzmann_equation}), we can numerically solve the time evolution of the heavy quark phase distribution $f(\vec{X},\vec{P},t)$. In the remaining part of this section, for conciseness, we will use the lower-case $\vec{x}$ ($\vec{p}$) to denote the 3-position (momentum) of the heavy quark.
  
The momentum distribution $f_{P}(\vec{p},t)$ of the heavy quark follows from its phase space distribution $f(\vec{x},\vec{p},t)$ as, 
\begin{equation}
    f_{P}(\vec{p},t)=\frac{\int f(\vec{x},\vec{p},t) d\vec{x}}{\int f(\vec{x},\vec{p},t) d\vec{x}d\vec{p}}.
\end{equation}
In our numerical calculations, we adopt the following Gaussian distribution as the momentum distribution at the initial time $t=0$,
\begin{equation}
f_{P}(\vec{p},t=0)=\frac{1}{(2\pi)^{0.5} \sigma} \exp{\left[ \frac{-(\vec{p}-\vec{p}_{0})^{2}}{2 \sigma^{2}}\right] },
\end{equation}
where $\vec{p}_{0}$ represents the momentum at which $f_{P}(\vec{p},t=0)$ reaches its peak value, and $\sigma$ is the width of the initial momentum distribution. For BE, we additionally assume a uniform distribution in the coordinate space at $t=0$.

Next, we compare the average squared momentum $\langle 
 p^{2}(t)\rangle $ and momentum distributions $f_{P}(\vec{p},t)$ obtained from the SSE and BE. For given momentum distribution, $\langle p^{2}(t)\rangle $ is defined as,
\begin{equation}
    \langle p^{2}(t)\rangle =\int |\vec{p}|^{2} f_{P}(\vec{p},t) d\vec{p}.
\label{p_square_integrate}
\end{equation}

In order to minimize the impact from the truncation artifacts in momentum space at $\vec{p}=\pm \pi\,\mathrm{GeV}$, we restrict the integration interval in Eq.~(\ref{p_square_integrate}) to $\left[-\frac{\pi}{3}\,\mathrm{GeV},\frac{\pi}{3}\,\mathrm{GeV}\right]$. The expected equilibrium momentum distribution of the heavy quark follows the uniform distribution, cf. Eq.~(\ref{equilibrium_relation}),  so the corresponding ensemble-averaged momentum squared is $\langle p^2\rangle _{\mathrm{eq}}=0.37\,\mathrm{GeV^{2}}.$

\begin{figure}[htbp]
    \includegraphics[scale=0.12]{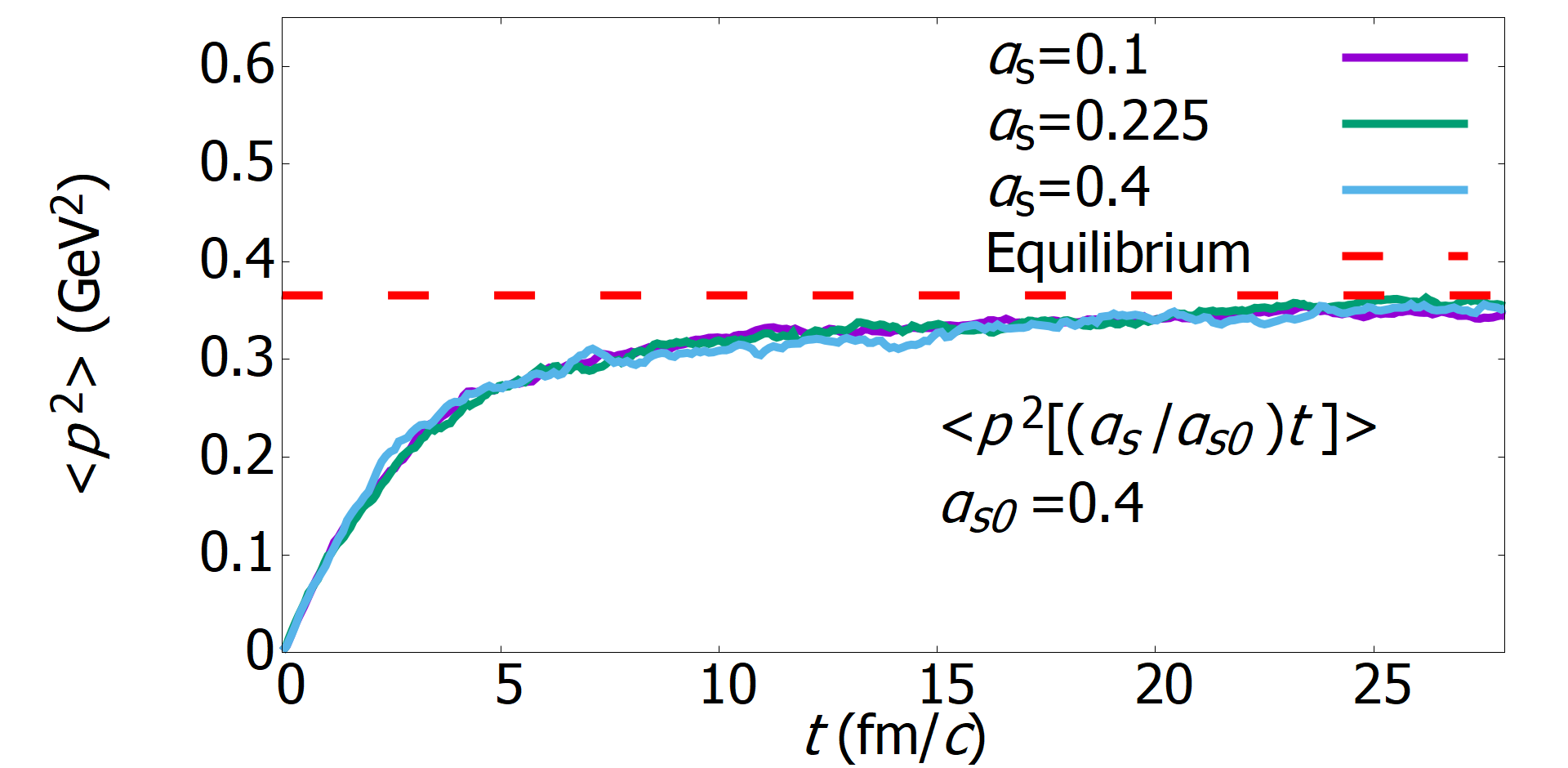}
    \caption{(Color online) The time evolution of $\langle p^{2}(t)\rangle $ from the SSE with different coupling constants between the heavy quark and the gluon field. The evolution time is rescaled with the factor of $\alpha_s/\alpha_{s0}$. The initial momentum distribution at $t$=0  is centered around $\vec{p}_0=0$, see text for the detail. The dashed line indicates the equilibrium value $\langle p^{2}\rangle _{\mathrm{eq}}=0.37\,\mathrm{GeV}^{2}$.}
    \label{fig:t1_diff_coupling}
\end{figure}

\begin{figure}[htbp]
  \includegraphics[scale=0.126]{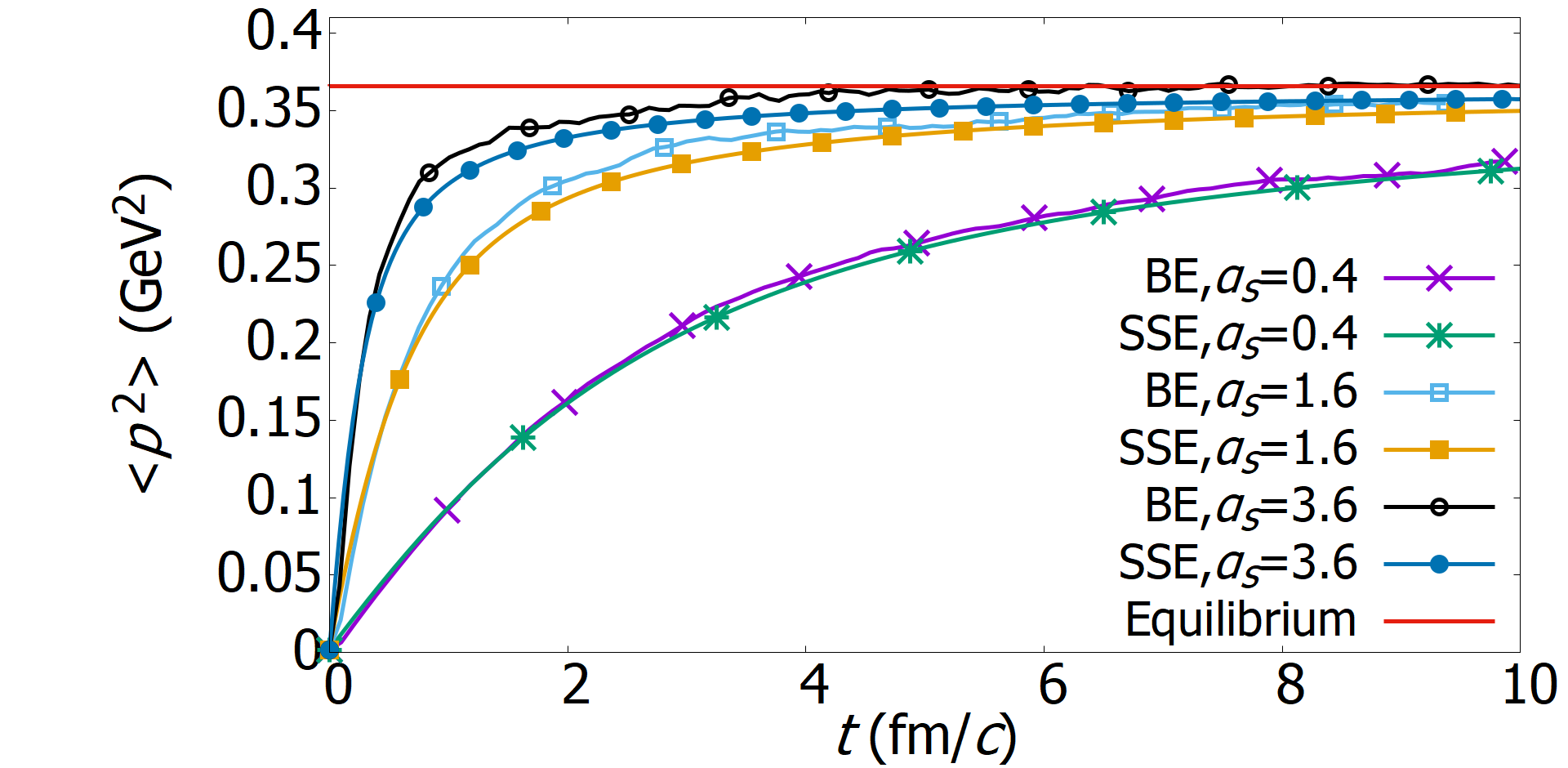} 
  \caption{(Color online) Comparison of the time evolution of $\langle p^{2}(t)\rangle $ from the SSE in Eq.~(\ref{sto_sch_equ}) and the Boltzmann equation (BE) in Eq.~(\ref{boltzmann_equation}) with $\alpha_{s}=0.4, 1.6, 3.6$. The dashed line indicates the equilibrium value $\langle p^{2}\rangle _{\mathrm{eq}}=0.37\,\mathrm{GeV^{2}}$. The initial momentum distribution at $t$=0  is centered around $\vec{p}_0=0$, see text for the detail.}
  \label{t1_I1.}
  \end{figure}

We compare the time evolution of $\langle p^{2}\rangle $ with $\vec{p}_{0}=0$ and $\sigma=0.043\,\mathrm{GeV}$ in Fig.~(\ref{fig:t1_diff_coupling}) and (\ref{t1_I1.}). As time evolves, $\langle p^{2}\rangle $ increases from zero toward the equilibrium value $\langle p^{2}\rangle _{\mathrm{eq}}=0.37\,\mathrm{GeV}^{2}$. From the Boltzmann equation in Eq.~(\ref{boltzmann_equation}), we can see that in the weak coupling limit, the scattering cross section is proportional to $\alpha_s$ if the gluon mass $m_{g}$ does not depend on the coupling constant. In Fig.~(\ref{fig:t1_diff_coupling}), we plot the evolutions of $\langle p^{2}[(\alpha_{s}/\alpha_{s0})t]\rangle $ with three different coupling constants $\alpha_{s}=0.1,0.225$ and $0.4$. Here, we rescale the evolution time by the ratio of the coupling constant, $\alpha_s/\alpha_{s0}$, where $\alpha_{s0}$ is the reference coupling constant appearing in the thermal mass of the gluon. We find that the three curves almost overlap with each other; that is, the evolution rate of $\langle p^{2}(t)\rangle $ is approximately proportional to the ratio of $\alpha_{s}/\alpha_{s0}$. These results indicate that the weak coupling constant condition is approximately satisfied with $\alpha_{s}\leq 0.4$.

In Fig.~(\ref{t1_I1.}), we compare the time evolution of $\langle p^{2}\rangle $ from the SSE with that from BE.  As $\alpha_{s}$ increases, the evolution rate calculated from the SSE gradually slows down compared to that from BE. This trend suggests that when $\alpha_{s}$ becomes large, nonperturbative effects may start to emerge during the time evolution. In Fig.~(\ref{t1_I1.}), all the results are averaged over 1000 events to further reduce the thermal fluctuations. 
  
\begin{figure}[!h]
  \includegraphics[scale=0.12]{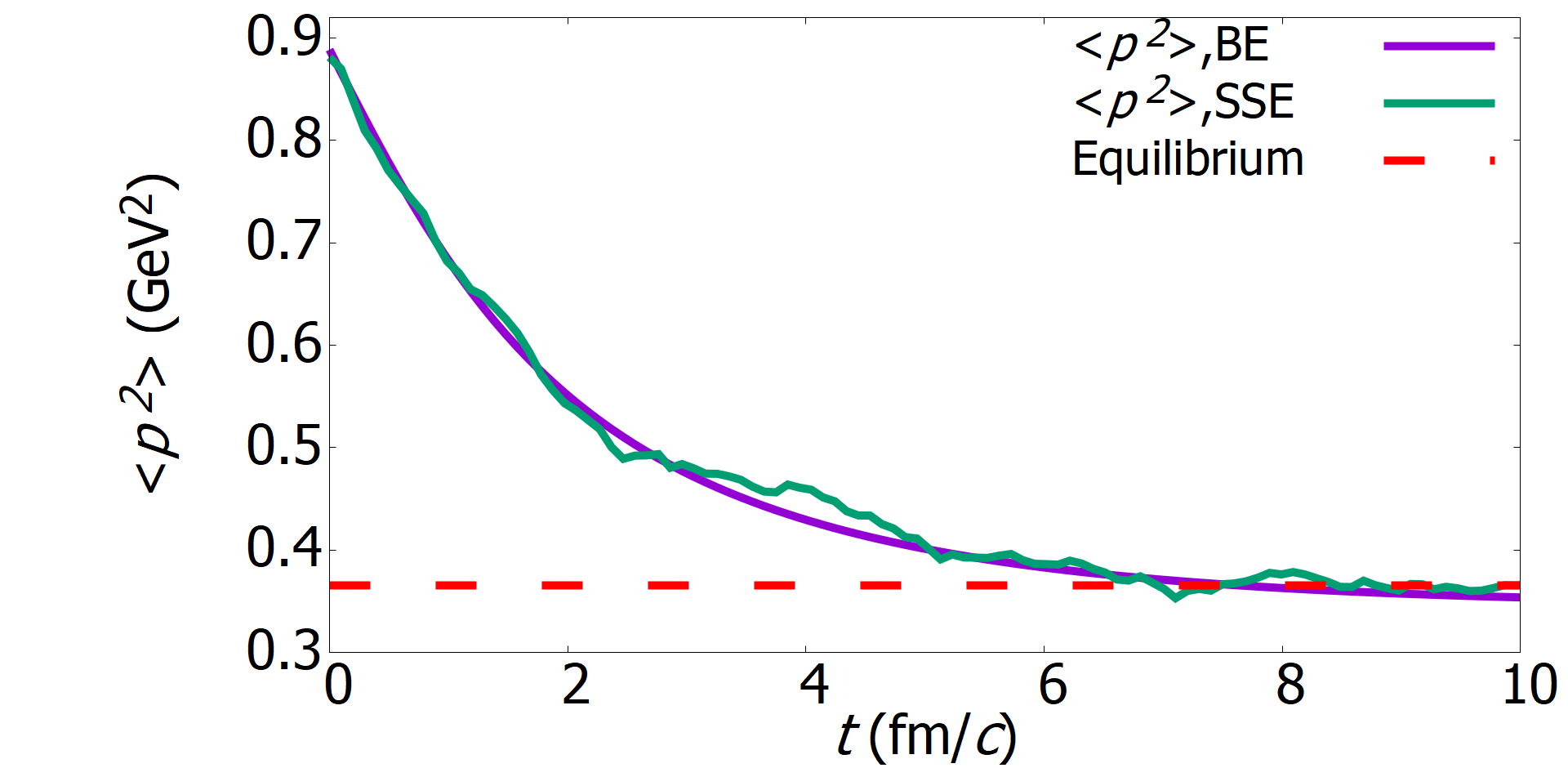} 
  \caption{(Color line) Time evolution of $\langle p^{2}(t)\rangle $ from the Boltzmann equation and SSE. The initial momentum distribution is peaked at $\vec{p}_{0}=0.2\pi\,\mathrm{GeV}$. Purple line: result from the Boltzmann equation. Green line: result from the SSE. The dashed line indicates the equilibrium value $\langle p^{2}\rangle _{\mathrm{eq}}=0.37\,\mathrm{GeV}^{2}$.}
  \label{t1_I1.8}
  \end{figure}

  In Fig.~(\ref{t1_I1.8}), we compare the time evolution of $\langle p^{2}\rangle $ from an initial momentum distribution centered around $\vec{p}_{0}=0.2\pi\, \mathrm{GeV}$ at $\alpha_s=0.4$. The resulting $\langle p^{2}\rangle $ decreases with time and approaches the equilibrium value of $\langle p^{2}\rangle _{\mathrm{eq}}=0.37\,\mathrm{GeV}^{2}$. The overlapping $\langle p^{2}\rangle $ evolution trends confirm that the consistency between the BE and SSE in the weak coupling condition is independent of the initial momentum distribution of the heavy quark.

\begin{figure}[htbp]
\subfigure{
  \includegraphics[scale=0.12]{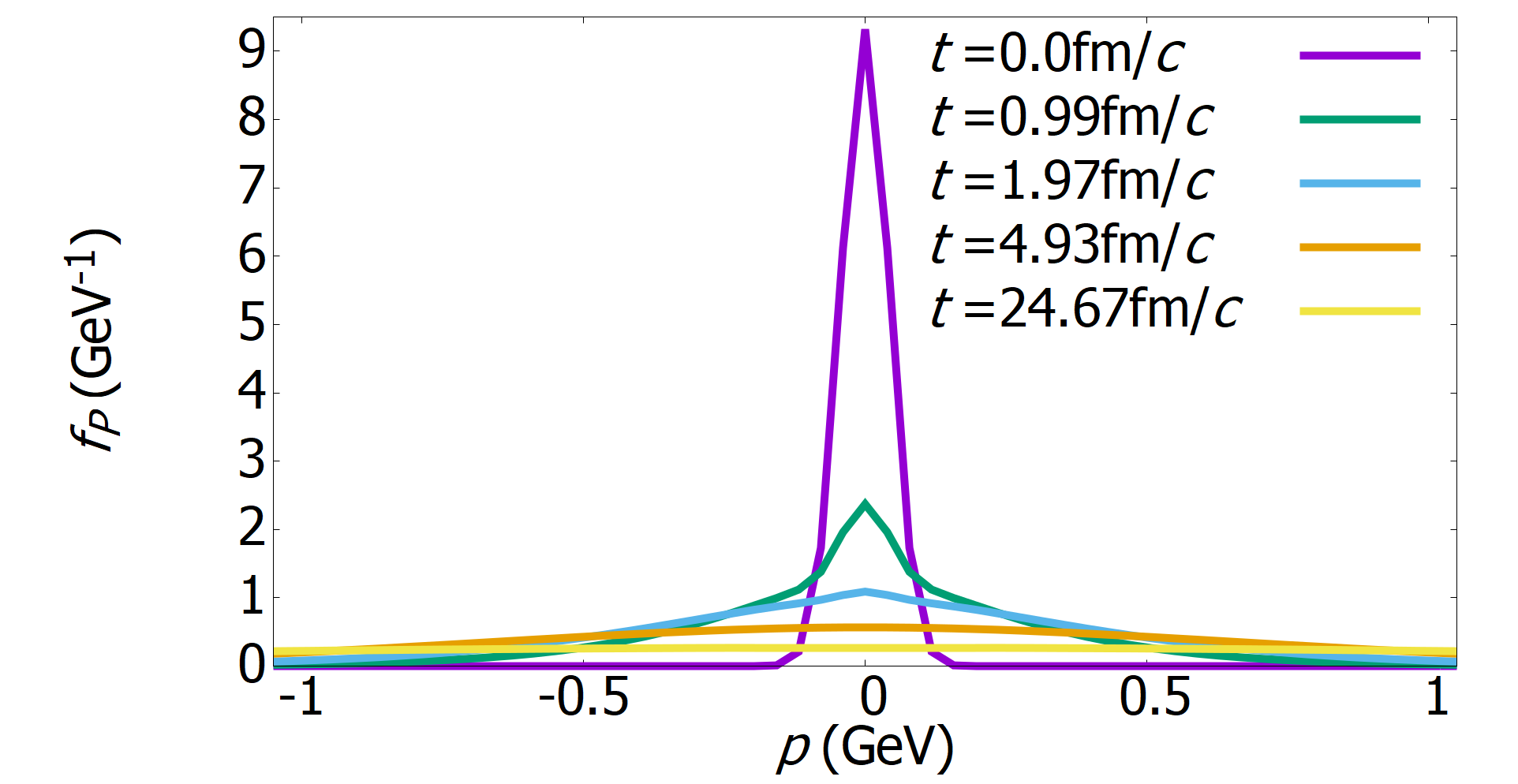} 
  }

  \subfigure{
  \includegraphics[scale=0.12]{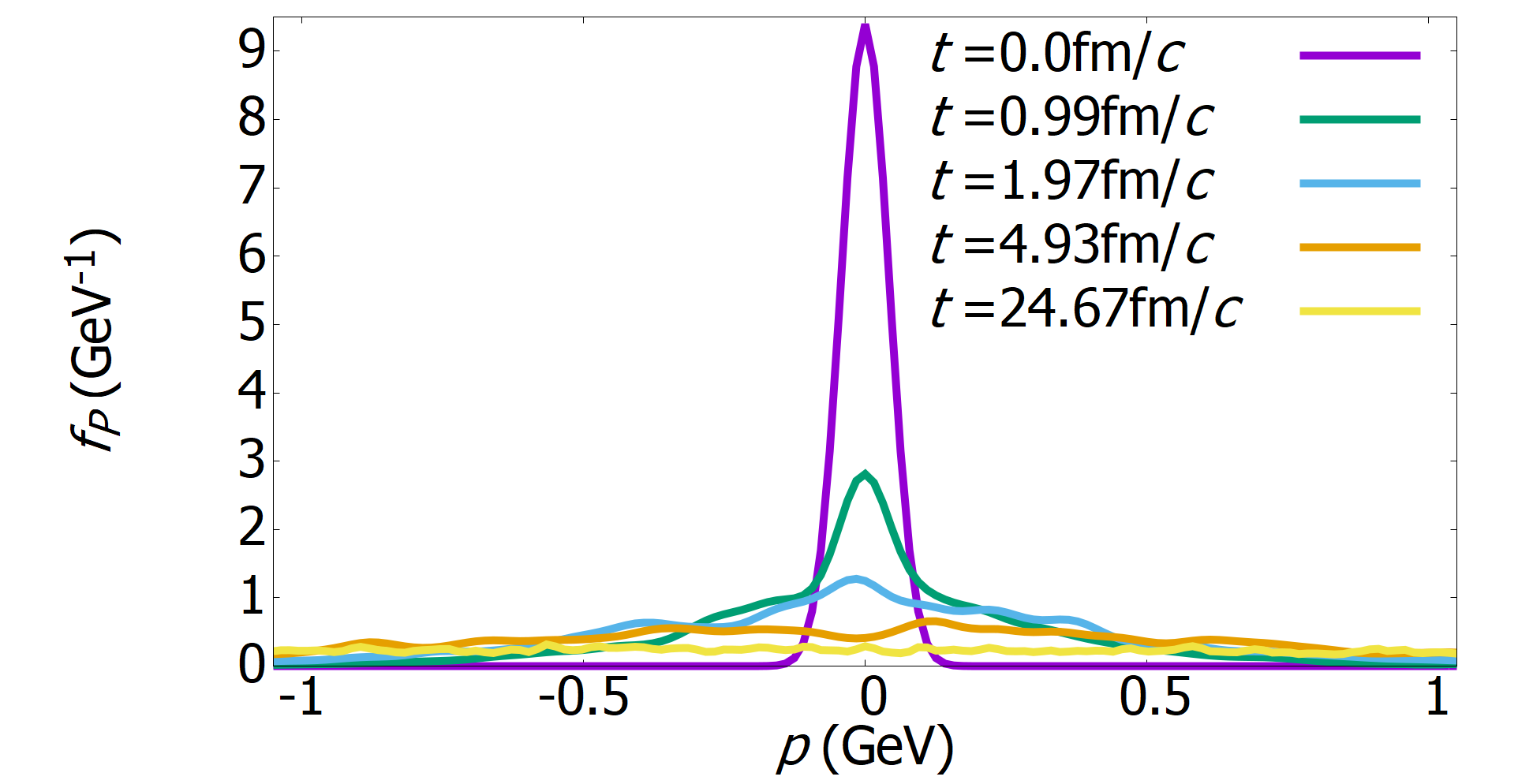} 
  }
  \caption{(Color online) Time evolution of the momentum distribution $f_{P}(\vec{p},t)$ of the heavy quark from the Boltzmann equation and SSE. The peak of the initial distribution is located at $\vec{p}_{0}=0$, see text for the detail. Top panel: results from the Boltzmann equation. Bottom panel: results from the SSE.}
  \label{distribution_evolution_I1.}
  \end{figure}

\begin{figure}[!h]
      \subfigure{
  \includegraphics[scale=0.12]{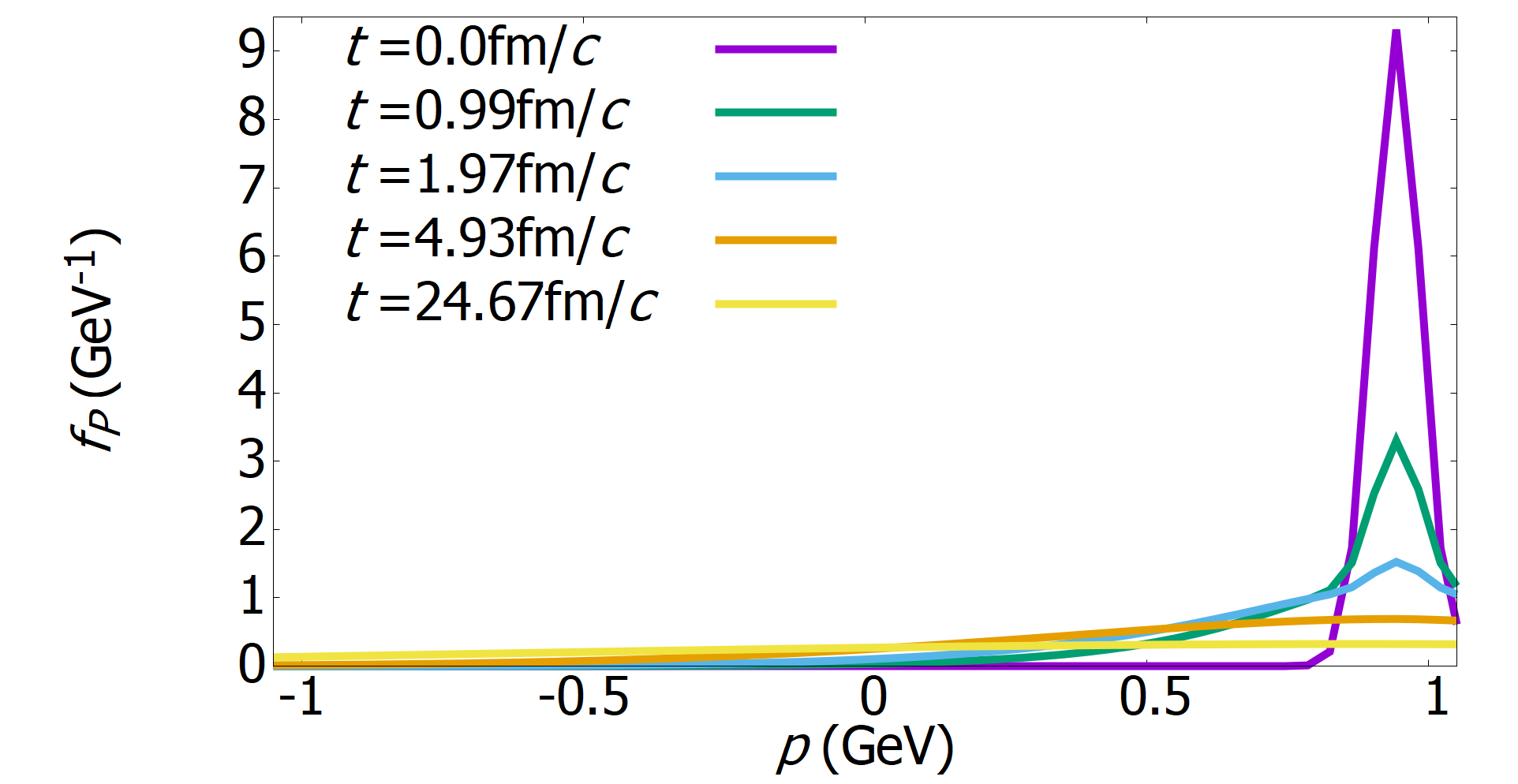} 
  }

  \subfigure{
  \includegraphics[scale=0.12]{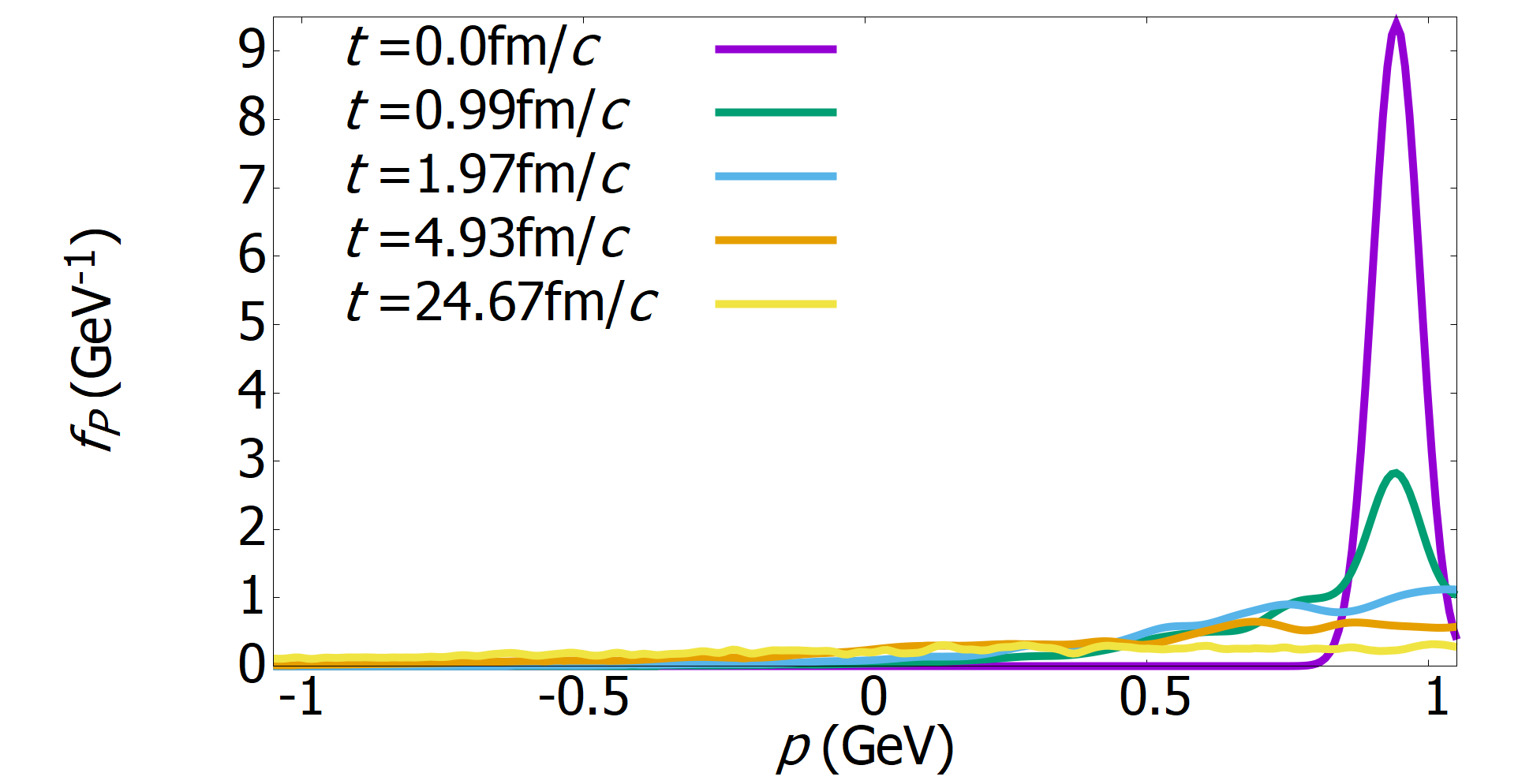} 
  }
  \caption{(Color online) Time evolution of the heavy quark momentum distribution $f_{P}(\vec{p},t)$ from the Boltzmann equation and SSE. The peak of the initial momentum distribution is located at $\vec{p}_{0}=0.2\pi\,\mathrm{GeV}$, see text for the detail. Top panel: results from the Boltzmann equation. Bottom panel: results from the SSE.}
  \label{distribution_evolution_I1.8}
  \end{figure}

The time evolution of the heavy quark momentum distribution $f_{P}(\vec{p},t)$ from the SSE and BE is compared in Fig.~(\ref{distribution_evolution_I1.}) (with $\vec{p}_{0}=0$) and Fig.~(\ref{distribution_evolution_I1.8}) (with $\vec{p}_{0}=0.2\pi\,\mathrm{GeV}$). The strong coupling constant used in this calculation is  $\alpha_s=0.4$. In both cases, the width of momentum distributions from the BE and SSE increases with the time $t$ and the momentum distributions approach the uniform distribution as $t$ increases, which is consistent with the Boltzmann's $H$-theorem, cf. Eq.~(\ref{equilibrium_relation}). One can see that for both initial conditions, the momentum distributions from the BE and SSE are close to each other at each intermediate time, which further confirms the consistency between the BE and SSE approach in the weak coupling condition..
    

\section{Conclusion}

In this study, we investigate the relationship between the Boltzmann equation (BE) approach and the recently constructed stochastic Schr\"{o}dinger equation (SSE) approach in the Keldysh Green's function framework. We find that the SSE approach describes the scattering processes between the heavy quark and the off-shell gluons in the thermal medium. The time scale for the phase rotation of the background gluon field is inversely proportional to the energy dispersion of the gluons in the thermal medium. When the coupling constant between the heavy quark and the gluon field is small, this process can be equivalently described by the BE approach with the scattering term obtained from the leading-order perturbative expansion of the heavy quark self-energy in the thermal medium.

Furthermore, we perform the numerical calculation of the time evolution of the heavy quark momentum distribution in both the SSE and BE approaches. The resulting momentum distributions are consistent with each other in the weak coupling limit, which confirms the connection between the SSE and BE found in the Keldysh Green's function framework. As the coupling increases, the results from the BE and SSE start to deviate, which signals the possible emergence of nonperturbative effects.

Compared to BE, the SSE approach has several advantages: First, in the SSE the heavy quark system evolves on the amplitude level and therefore this approach can potentially capture a more complete set of quantum effects, such as the (de)excitation of bound states through the interaction with the medium. Secondly, the SSE provides a straightforward nonperturbative framework to study the time evolution of strong coupling systems in the thermal medium. 

The application of the SSE can be extended in several directions: First, we plan to apply the SSE to bound state systems consisting of heavy quarks, such as the heavy quarkonium systems, and study their formation and dissociation in the thermal medium. Secondly, we can implement more realistic  background fields to simulate the thermal medium. For example, the transverse components of the gluon field as well as the time dependence and the flow effects of the thermal medium can be considered. Finally, besides the gluon field, we can include time-dependent electromagnetic fields in the SSE  and study their effects on the evolution of the heavy quark systems in the thermal medium.

\section{Acknowledgment}
We thank Baoyi Chen and Min He for useful discussions. X. Z. is supported by new faculty startup funding by the Institute of Modern Physics, Chinese Academy of Sciences, by Key Research Program of Frontier Sciences, Chinese Academy of Sciences, Grant No. ZDBS-LY-7020, by the Foundation for Key Talents of Gansu Province, by the Central Funds Guiding the Local Science and Technology Development of Gansu Province, Grant No. 22ZY1QA006, by Gansu International Collaboration and Talents Recruitment Base of Particle Physics (2023-2027), by International Partnership Program of the Chinese Academy of Sciences, Grant No. 016GJHZ2022103FN, by National Natural Science Foundation of China, Grant No. 12375143, by National Key R\&D Program of China, Grant No. 2023YFA1606903 and by the Strategic Priority Research Program of the Chinese Academy of Sciences, Grant No. XDB34000000. 

\bibliography{biblio}

\end{document}